\documentclass[fleqn,usenatbib]{mnras}

\usepackage{amsmath} 
\usepackage{fourier}

\usepackage[english]{babel}															
\usepackage[protrusion=true,expansion=true]{microtype}	

\usepackage[pdftex]{graphicx}	
\usepackage{subcaption}
\usepackage{url}
\usepackage[thinc]{esdiff}
\usepackage{physics}
\usepackage[normalem]{ulem}

\usepackage[T1]{fontenc}

\DeclareRobustCommand{\VAN}[3]{#2}
\let\VANthebibliography\thebibliography
\def\thebibliography{\DeclareRobustCommand{\VAN}[3]{##3}\VANthebibliography}

\newcommand{\prodimo}{\textsc{ProDiMo} }

\usepackage{graphicx}	
\usepackage{amsmath}	
\usepackage{amssymb}	

\title[]{Impacts of Energetic Particles from T Tauri Flares on Inner Protoplanetary Discs}

\author[V. Brunn et al.]{
V. Brunn,$^{1}$\thanks{e-mail: valentin.brunn@umontpellier.fr}
Ch. Rab,$^{2,3}$
A. Marcowith,$^{1}$
C. Sauty,$^{1,4}$
M. Padovani,$^{5}$
and C. Meskini$^{1}$
\\
$^{1}$Laboratoire Univers et Particules de Montpellier, Universit\'e de Montpellier/CNRS, place E. Bataillon, cc072, 34095 Montpellier, France\\
$^{2}$Universit\"ats-Sternwarte, Fakult\"at f\"ur Physik,   Ludwig-Maximilians-Universit\"at M\"unchen, Scheinerstr.~1, 81679 M\"unchen, Germany\\
$^{3}$ Max-Planck-Institut für extraterrestrische Physik, Giessenbachstrasse 1, 85748 Garching, Germany\\
$^{4}$Laboratoire Univers et Th\'eories, Observatoire de Paris, Université PSL, Universit\'e Paris Cit\'e, CNRS, F-92190 Meudon, France\\
$^{5}$INAF-Osservatorio Astrofisico di Arcetri, Largo E. Fermi 5, 50125 Firenze, Italy\\
}

\date{Accepted XXX. Received YYY; in original form ZZZ}

\pubyear{2024}

\begin{document}
\label{firstpage}
\pagerange{\pageref{firstpage}--\pageref{lastpage}}
\maketitle

\begin{abstract}
T Tauri stars are known to be magnetically active stars subject to strong flares observed in X-rays. These flares are likely due to intense magnetic reconnection events during which a part of the stored magnetic energy is converted into kinetic energy of supra-thermal particles. Since T Tauri stars are surrounded by an accretion disc, these particles may influence the disc dynamics and chemistry. This work continues on a previous stationary model, which showed that energetic particles accelerated during flares can produce a strong ionisation rate at high column densities in the inner accretion disc. The present model includes non-stationary sequences of flaring events sampled by a Chandra X-ray survey of nearby young stellar objects. We calculate the averaged ionisation rate expected in a radius range from 0.08 to 0.6 au from the central star. We confirm that energetic particles produced by the flares dominate the ionisation of the disc up to column densities of $10^{25}~\rm{cm^{-2}}$. We further study the main consequences of this additional source of ionisation on the viscosity, the accretion rate, the volumetric heating rate and the chemical complexity of inner protoplanetary discs.
\end{abstract}




\section{Introduction}

T Tauri stars are young stars, typically of a few million years old, still contracting and accreting material from their surrounding circumstellar discs. The way T Tauri stars accrete the surrounding infalling matter is still an open subject in astrophysics \citep{2014A&A...570A..82V,hartmann2016accretion,2016A&A...591L...3M}. All mechanisms that describe accretion-ejection processes involve the interaction between magnetic fields and ionised disc matter (e.g., \citealt{Shakura73,blandford1982hydromagnetic}). 

The magnetic field geometry of these stars is much more complex than those of non-accreting cool stars \citep{donati2007magnetic, gregory2008non}. Some field lines might become twisted due to the differential rotation between the disc and the star, causing them to bulge outwards and possibly even eject matter \citep{zanni2013mhd}. This process possibly leads to very strong flaring events observed in X-rays \citep{Feigelson99,getman2008a,Getman_2021a}. It is thought to be caused by magnetic reconnection events. Magnetic reconnection transforms a fraction of the system magnetic energy into particle kinetic energy, producing bursts of high-energy particles and radiation as observed in our Sun \citep{2011LRSP....8....6S,2017LRSP...14....2B,2018SSRv..214...82O}. T Tauri flares appear to be thousands of times more energetic than solar flares \citep{Getman_2021a}. They are expected to produce particle fluxes that are five orders of magnitude higher than those produced by the Sun \citep{Feigelson2002,Rab17}. The high-energy particle flux produced by T Tauri stars strongly impacts the surrounding environment \citep{Rodgers-Lee17,Rab17,2019ApJ...883..121O,waterfall2020predicting, brunn2023ionization}. In particular, these particles can ionise the gas in the surrounding disc \citep{brunn2023ionization}.
Disc ionisation is a critical process in T Tauri discs, as it couples the disc material to magnetic fields, which affects the dynamics and chemistry of the whole disc \citep{2013ApJ...772....5C,2019ApJ...883..121O,2022ApJ...928...46W} and jets \citep{sauty2019jet, Jacquemin-Ideetal19, Ray21}. Magnetic fields can also trigger instabilities in the gas that can drive accretion of disc material onto the star \citep{hartmann2016accretion,2021JPlPh..87a2001P}. Besides, ionisation drives chemistry in the disc, leading to the formation of molecules and complex organic compounds like amino acids \citep{2004A&A...417...93S}, which are the building blocks of prebiotic chemistry. Ionisation also plays a role in the heating and cooling of discs \citep{Glassgold_2012}, a process known to be essential in the launching of jets \citep{Ray21}. Understanding the mechanisms that drive disc ionisation appears then necessary for understanding the formation and evolution of stars and the dynamics of accretion-ejection processes.

In a first article (\citealt{brunn2023ionization}, hereafter B23), we studied in stationary framework, the ionisation rate produced by magnetic reconnection events occurring above the inner disc. We call inner disc the region at a radial distance between $R=0.08$ au and $0.6$ au from the central star. We estimated ionisation rates produced by protons, electrons, and secondary electrons accelerated from these flares. Our model indicates that in the close environment of the flare, ionisation by non-thermal particles is the dominant ionisation source in the inner disc of T Tauri stars. It produces ionisation rates several orders of magnitude higher than previously proposed sources such as X-rays, galactic Cosmic Rays (GCRs), and radionuclides. B23 also proposed a comparative analysis of parameters to determine the conditions under which ionisation rates produced by magnetic reconnection are dominant. The results suggest that energetic particles (EPs) accelerated above the disc during magnetic reconnection events enhances the ionisation rates in the inner disc region of T Tauri stars. They are a new source of ionisation important to be included in magneto-hydrodynamics (MHD) or astrochemical codes.

Studying ionisation in protoplanetary discs with a non-stationary approach allows to calculate the mean ionisation fraction in the disc. Ionisation is a source of heat and a precise computation is crucial for understanding how jets are launched from the disc. Such results will be of interest for MHD simulations studying jet launching to constrain the heating rate based on physical constrains. Another important aspect is to couple the ionisation models with non-ideal MHD simulations of protoplanetary discs. This will provide insights into how the magnetic fields interact with the material in the disc and how it may shape the system and affect its evolution. Another goal of this work is to improve the astrochemical codes of protoplanetary discs. By developing more accurate models of the ionisation processes at play in these systems, we shall have a better understanding of the disc chemical evolution leading to the formation planetesimals and the synthesisation of the building blocks of live. 

In this paper we study in detail the non-thermal particle injection in the star-disc environment. It is the first of a series in which we shall study ionisation of T Tauri discs due to flares in a non-stationary framework. In this paper, we focus on the effects of the flares occurring close to the star, in the inner disc. For this study, we benefit from the X-ray observations of Chandra presented by \citet{Getman_2021a} and \citet{Getman_2021b}.

The layout of the paper is as follows. In section \ref{S:METHOD} we summarise the stationary model as presented in B23, then present the physical model of flares, the chemical model and the model to derive the viscous $\alpha$ parameter in the disc from the ionisation fraction. In section \ref{S:RESULTS}, we present our results for the case of one flare, then we extend them to the time-dependent solutions from our multiple flare model. We consider the impact of non-thermal particle ionisation over disc chemistry, the development of the magneto-rotational instability and the heat deposition in the disc and at the base of the jet. In section \ref{S:DISCUSSION}, we propose a general discussion about theoretical perspectives of our work and its limitations before concluding in Sec. \ref{S:CONCLUSION}.

\section{Method}\label{S:METHOD}\subsection{Summary of the stationary model in B23}

Elaborating on the work by \citet{2011MNRAS.415.3380O} and \citet{2019A&A...624A..50C}, B23 propose an uncovered source of ionisation in the inner discs of T Tauri stars, the ionisation by flares triggered by magnetic reconnection. The idea was also inspired by the results of the simulation of magnetospheric ejection by \citet{zanni2013mhd} where the typical magnetic field topology of magnetic reconnection can clearly be identified in the interaction zone of the magnetic field of the star and the magnetic field of the disc. On the other hand, it has been routinely observed in the Sun and in laboratory experiments that such magnetic reconnection events are very efficient at accelerating particles to energies $\lesssim 1$ GeV. Given the geometry of the field lines above the disc, we expect that this geometry favor the propagation of EPs from the acceleration region down to the disc. 

Once these particles penetrate into the disc, they suffer from energy losses. At energies $\lesssim 1$ GeV, particles mainly lose their energy through ionisation losses. In other words, they are very efficient at ionising the disc. This supports the idea that our model could be a significant source of ionisation in the inner disc. B23 estimate the ionisation rate produced by EPs accelerated by magnetic reconnection events in the region of star-disc magnetic field interaction and proposes a parametric study. The methodology of the B23 model is as follows. 

First, B23 estimate the flux of supra-thermal particles produced by a flare. This estimation is based on X-ray observations from the COUP (Chandra Orion Ultradeep Project) point source catalogue \citet{2005ApJS..160..469F, getman2008a}. Using these data, B23 detail how to estimate the flare size and the electron density of the flare based on a the observed flare temperature. B23 then develop a model of particle acceleration which, allows to estimate the non-thermal particle flux from the temperature, thermal electron density and size associated to the flare. The acceleration model also relies on the results of numerical simulations of magnetic reconnection to estimate the power-law index of the flux of non-thermal particles. Thus, B23 propose an estimation of the non-thermal particle flux produced by flares of T Tauri star based on micro-physical considerations. This flux takes as parameter the flare temperature, the index of the EPs distribution and the maximum energy of the EPs distribution. 

Subsequently, B23 describe their model for particle transport. This transport model allows for the estimation of the propagated particle flux within the disc based on the initial particle flux produced by magnetic reconnection, as a function of column density. Particle transport is described within the Continuous Slowing Down Approximation (CSDA), which is applicable under the assumption that the pitch angle of particles remains constant and that the energy losses due to collision with the medium are small compared to the kinetic energy of the particle. B23 computed the propagated flux for column densities ranging from $10^{19}$ cm$^{-2}$ to $10^{25}$ cm$^{-2}$ in order to treat the propagation of particles in the CSDA regime \citep{Padovani18}. Within this approximation, given that the gyroradius of particles with energy $E<1$ GeV is small compared to the characteristic size of the disc, the trajectories of the particles are determined by the large-scale magnetic field lines. B23 provides a detailed description of the energy loss processes using the formalism of the energy loss function. The loss function is defined as the energy loss per unit column density and depends both on the incoming particle species (protons or electrons) and the composition of the target medium. 

The composition of the target medium, i.e. the abundance of various chemical species within the disc through which the particles travel, is determined using the radiation thermo-chemical code \prodimo (see Sect. \ref{sec:Prodimo} for a further description). B23 focuses exclusively on the losses associated with the dominant species: free electrons, H, H$_2$ and He. We have estimated that the energetic losses due to more complex species accounts for less than 5\% of the total energy losses. Consequently, we neglect the contribution of metals and more complex molecules to the energetic losses in order to simplify the transport model. 

In B23, the electron loss function is deduced from the data provided by \citet{dalgarno1999electron} at low energy ($\rm <1keV$), which are based on theoretical calculations. At high energy ($ \rm >1keV$) B23 use the National Institute of Standards and Technology database\footnote{\url{https://physics.nist.gov/PhysRefData/Star/Text/ESTAR.html}\label{NIST}}, see \citet{2022A&A...658A.189P} for an updated description of the electron energy loss function. At low energy, energy losses in molecular gases are controlled by rotational and vibrational excitations, at intermediate energy by ionisation and at high energy by radiative losses. The data of the Stopping and Range of Ion in Matter \citep{ziegler2010srim} are used to construct the proton loss functions. At high energies, above the threshold $E^\pi = 280 \rm MeV$, we add
energy losses due to pion production, as reported by \citet{Padovani18}. At very low energy ($<1-10 \rm eV$) Coulomb losses are dominant, we use the expression from \citet{schlickeiser2013cosmic}. 

Whether the cross sections and loss functions are determined experimentally or theoretically, the uncertainties in their determination by the authors cited in B23 are negligible compared to the uncertainties linked to the parameters of the acceleration model. Additionally, B23 include the flux of secondary electrons produced issued from the ionisation of the medium by primary particles \citep{ivlev2015interstellar}. 

Finally, B23 show that, in a stationary configuration, EPs are a powerful source of local ionisation, with ionisation rates that exceed X-ray, stellar EPs, and radioactivity contributions in the inner disc by several orders of magnitude. For the reference case of a 1MK flare at 0.1 au propagating along a vertical magnetic field line, B23 find an ionisation rate $\zeta=10^{-9} ~\rm  s^{-1}$ at a column densities of $10^{25} ~\rm cm^{-2}$, while the ionisation rate at this depth due to GCR or X-rays produced in a $1~\rm MK$ stellar flare is of the order of $10^{-17} ~\rm  s^{-1}$, while it is of the order of $10^{-18} ~\rm  s^{-1}$ due to radionuclides. B23 then proceed to a parametric study where they consider various spectral indices for the EPs flux, various flare temperatures, different poloidal magnetic configurations as well as toroidal components along which the EPs can propagate, and different distance from the star where EPs are entering into the disc.

Although in conclusion B23 indicate that their assumptions may lead to overestimate the ionisation rate even if this process is shown to be a potentially dominant ionisation processes in the inner disc of T Tauri stars. As there are several parameters in the model that are difficult to constrain either experimentally or observationally, B23 conducted a comparative analysis of these parameters. The aim of this analysis is to define a range of flare parameters so that the ionisation rates produced are dominant over other ionisation sources. B23 anticipate that this will be the case for: (i) a reconnection process that accelerates particles following an injection flux with a power law  in kinetic energy $j \propto E^{-\delta}$ for $\delta < 6$, (ii) flares with temperatures above 1 MK, (iii) particles propagating along the field line with a ratio of the toroidal component to the poloidal component $b_{\rm g}=B_{\phi}/B_{\rm pol} < 1$.

In the present article, we have updated the method for calculating the non-thermal particle density based on temperature of B23 (see App. \ref{app:Nonthermaldensitynormalisation}). This revised approach is applied to determine the injected flux of non-thermal particles. We note that the ionisation rate is affected only by a factor of 2, which is relatively modest regarding the lack of constraints on the injected flux. This adjustment has insignificant implications for the chemistry involved.

\begin{figure}
    \centering
    \includegraphics[width=1\linewidth]{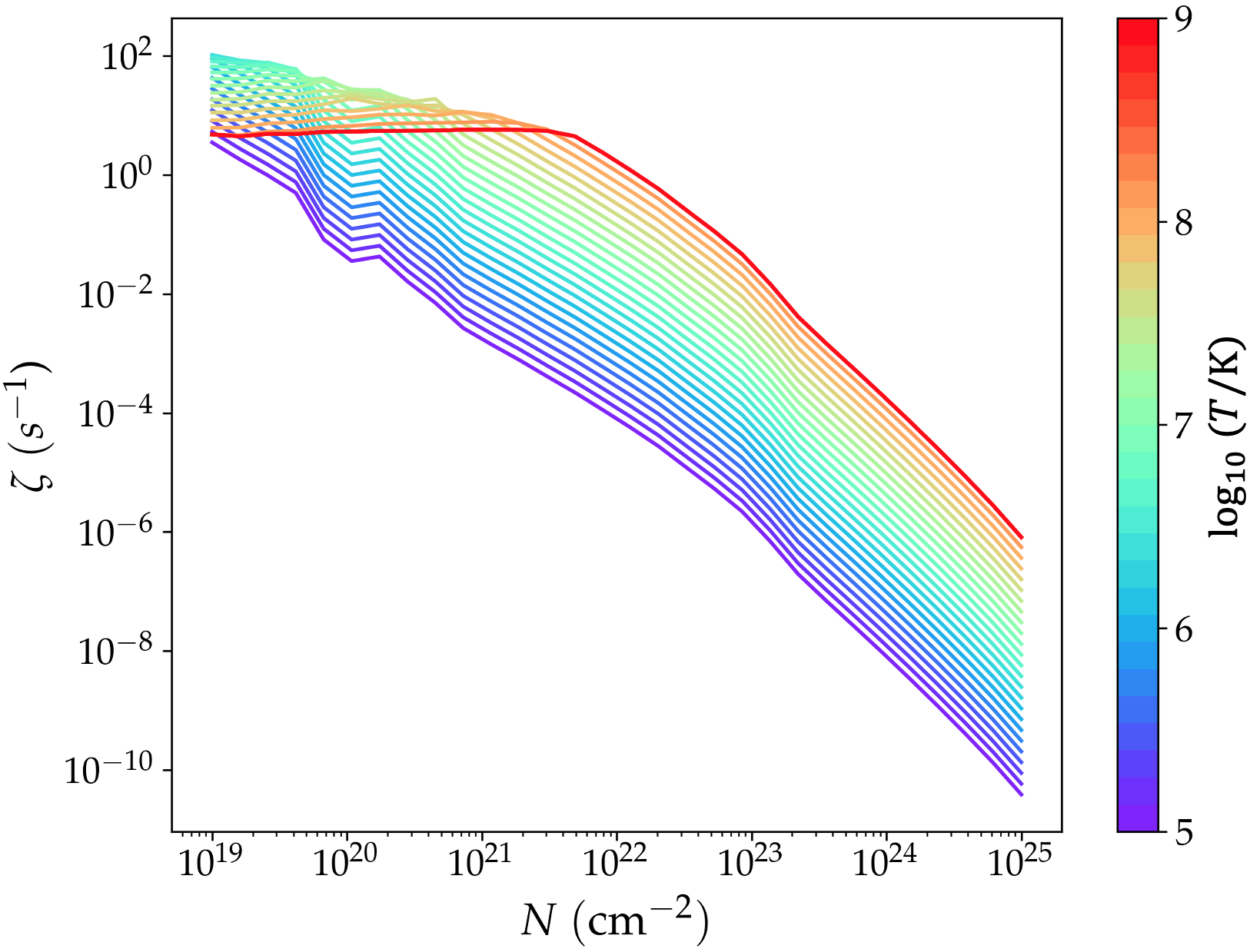}
    \caption{Non-thermal particle ionisation rate as a function of column density for a range of flare temperature between $10^5$ K and $10^9$ K. The results are obtained assuming the fiducial parameters as described in B23. The flare is assumed to occur at a radius of 0.1 au from the star.}
    \label{fig:IonizationRateTemperature}
\end{figure}

According to B23, the temperature strongly affects the ionisation rate in the disc as it determines both the normalisation and the injection energy of the non-thermal component of the particle flux. Figure \ref{fig:IonizationRateTemperature} shows the ionisation rates generated by flares with temperatures ranging from $10^5$ K to $10^9$ K in the fiducial configuration of B23. This configuration assumes a power-law for the energy distribution function $F(E) \propto E^{-\delta}$, and $\delta=3$ for particles produced by a magnetic reconnection event. It also assumes that magnetic field lines are perpendicular to the disc plane and that the EPs follow helical trajectories around the vertical lines with a homogeneous distribution of pitch angles. 

The assumption that the magnetic field is perpendicular to the disc plane stems from the results from simulations of \cite{mattiafendt2022,mattiafendt2020a,mattiafendt2020b,zanni2013mhd} where the authors take into account turbulence and dynamo. They show that, especially when turbulence is saturated, the magnetic field is mostly vertical. Simulations in \citet{jacqueminetal2021} of turbulent flows show similar results with almost vertical magnetic field lines. These results are comparable to what is obtained in self-similar accretion-ejection models \citep{ferreira1997}. However, self-similar models including magnetorotational instability \citep{jannaud2023} may show more complicated bent magnetic structures such as those we explored in B23. This however does not affect strongly the results. Of course, as the disc is axisymmetric, there should be an extra toroidal magnetic field that is acting as a guide field.
.

\subsection {Non-stationary flare model}\label{seq:FlareModeling}
As stated above, the model in B23 is stationary, the model estimates the ionisation rate below the flare, at its maximum luminosity. Consequently, the resulting ionisation rate is not representative of the ionisation of the whole disc on an extended period of time. This is why a time-dependent model should be considered. 

In this section, we propose a stochastic model to simulate the X-ray emission of a T Tauri star. Our aim is to calculate the resulting ionisation rate due to flares using the X-ray luminosity as an input. Therefore, it is needed to establish a relationship between the peak X-ray luminosity ($L_{X,\rm pk}$)  and the corresponding flare temperature $T_f$.

To proceed, we rely on observations from Chandra. The work of \citet{getman2008a} includes a table that displays the X-ray luminosity and the temperature of 216 flares in the COUP sample. Linear regressions are applied to the data. First, we deduced from this the relationship $\log L_{X,\rm pk} - \log T_{\rm obs}$ linking the peak X-ray luminosity in the Chandra detection energy band to the temperature deduced in this band $T_{\rm obs}$. Second, we deduce the relationship $\log T_f - \log T_{\rm obs}$, linking the bolometric and X-ray flare temperature. \citet{getman2008a} indicate that because the peak X-ray emission of flares with temperature $T_{\rm obs}>200$ MK is outside the Chandra sensitivity band, the measurements deduced for these flares are highly imprecise. So, to perform the regression, we only consider a restricted sample of 179 flares with $T_{\rm obs}<200$ MK. 
\begin{figure*}
\begin{subfigure}{.5\textwidth}
  \centering
  \includegraphics[width=1\linewidth]{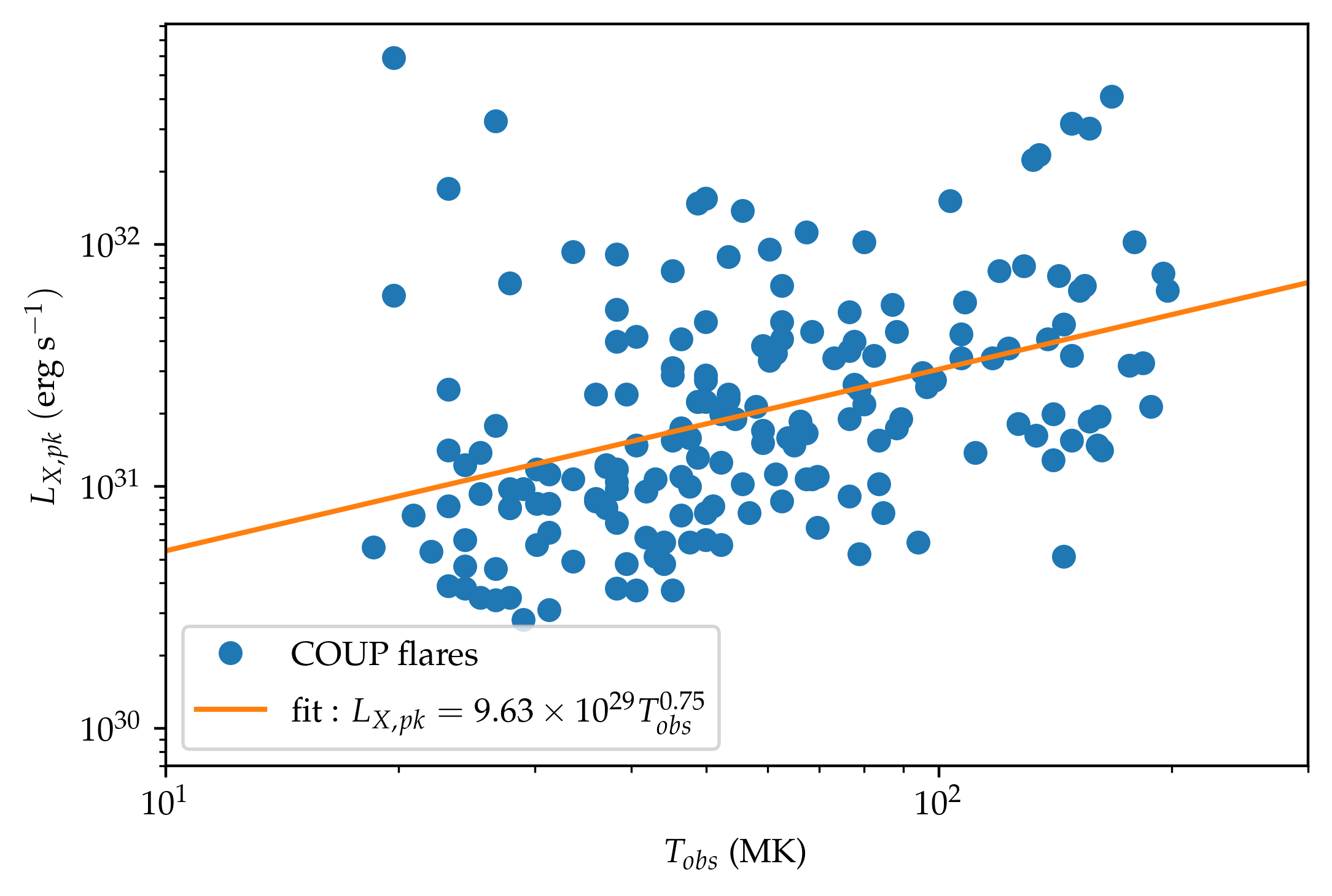}
  \caption{}
  \label{fig:LXTobsfit}
\end{subfigure}%
\begin{subfigure}{.5\textwidth}
  \centering
  \includegraphics[width=1\linewidth]{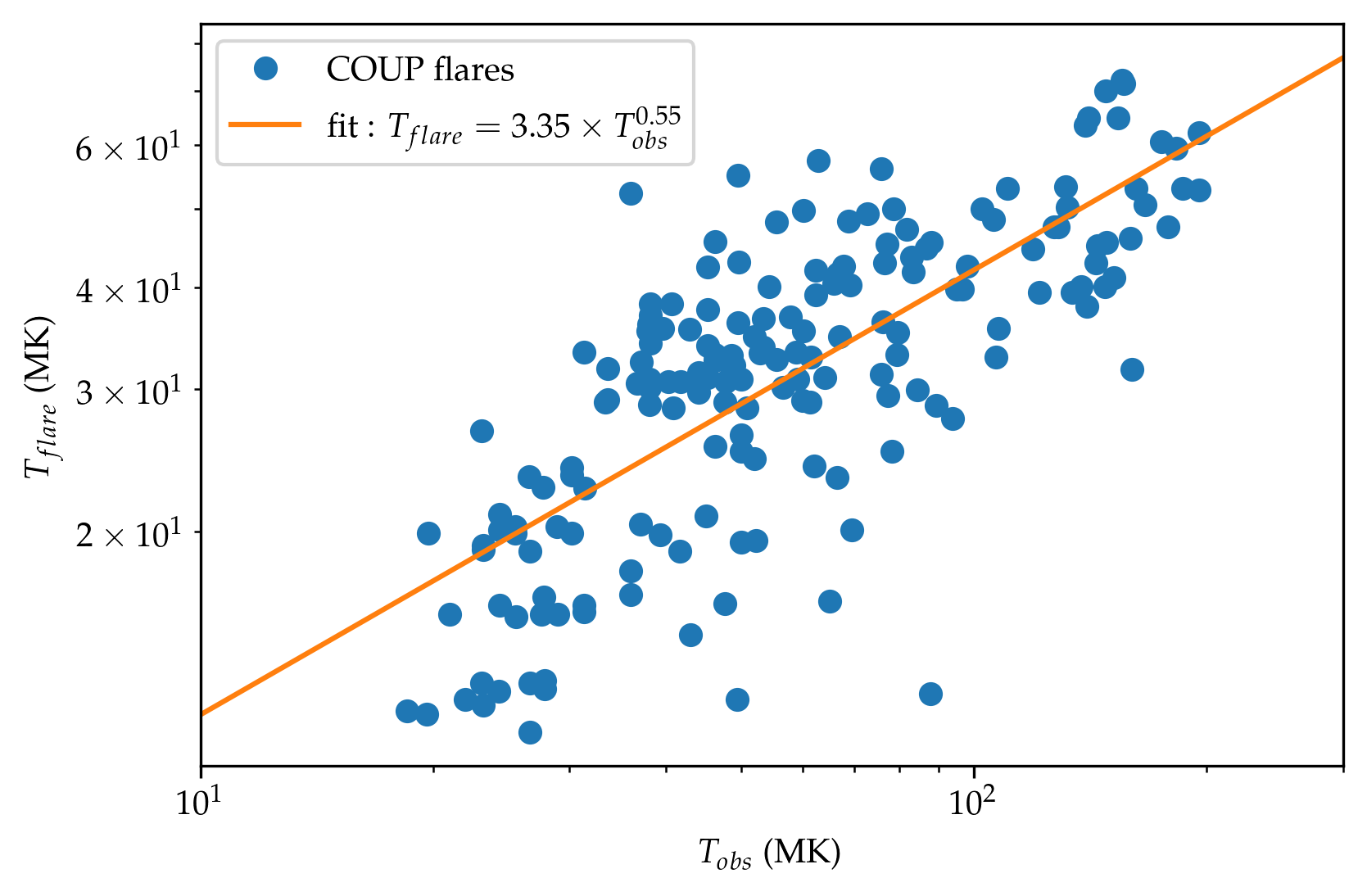}
  \caption{}
  \label{fig:TobsTflarefit}
\end{subfigure}
\caption{Panel (a) shows the flares peak X-ray luminosity $L_{X,\rm pk}$ dependence as function of the flare observed temperature $T_{\rm obs}$ in the Chandra band. Panel (b) shows the flares bolometric temperature $T_f$ dependence as function of the flare observed temperature $T_{\rm obs}$ in the Chandra band. The fits are plotted in solid orange line, while the data from \citet{getman2008a} are plotted as blue dots.}
\label{fig:TemperatureLumLinearReg}
\end{figure*}

In Fig. \ref{fig:TemperatureLumLinearReg}, we show the scattering plots of Chandra data points along with their corresponding linear regression. In Fig. \ref{fig:LXTobsfit}, we plot the relation between the peak X-ray luminosity and the observed temperature of the flare $L_{X,\rm pk}(\rm erg~ s^{-1})=9.63\times10^{29} \left(\frac{T_{\rm obs}}{\rm MK}\right)^{0.75}$. In Fig. \ref{fig:TobsTflarefit}, we plot the relation between the observed temperature in the Chandra band and the bolometric temperature of the flare $T_f=3.35\times T_{\rm obs}^{0.55}$ (both temperatures are in MK units). Combining these two equations, the relation between the observed X-ray luminosity and the temperature of the flare can be derived as,
\begin{equation}
    T_f= 3.45 \times \left(\frac{L_{X,\rm pk}}{10^{30} ~ \rm erg ~s^{-1}}\right)^{0.73} \quad \rm MK.
    \label{eq:TflareLX}
\end{equation}

We present now the stochastic model used for probing the parameter space relevant for T Tauri flares. Our main objective is to simulate the X-ray emission from young stars during a Keplerian period $P_K$. For solar mass stars, the Keplerian period at 0.1 au, $P_K$, is approximately 10 Ms ($\approx  0.3$ years).

Our methodology is based on a random sampling of flare distributions, based on three parameters, the waiting time between two events, the X-ray luminosity and the flare duration.
To randomly sample the distributions, we use an inverse function technique, as detailed in \citet[Chapter 6]{malzac:tel-00010420}.

A comprehensive observational analysis spanning six decades of solar activity, recorded by instruments such as ISEE-3/ICE, HXRBS/SMM, WATCH/GRANAT, BATSE/CGRO, and RHESSI \citep{Aschwanden_2010}, enable the identification of the distribution of the waiting times between solar flare events. We assume that we can extrapolate this solar flare temporal distribution to T Tauri star flares. This empirical distribution helps estimating the flare frequency within our designated period of study, while also giving a time of occurrence to each flare event. By relying on solar observations, we can estimate the frequency distribution of flares with luminosity that are lower than those observable on T Tauri stars.

Subsequently, to ascribe a luminosity value to every individual flare, we refer to the luminosity distribution of X-ray flares from young stars, as observed by Chandra \citep{Getman_2021a}. In addition, the data deduced from the X-ray flares of these young stellar objects provide the distribution of the flares duration.

\subsubsection{Waiting Times Distribution}\label{sec:waitingtimes}

Observations of waiting times between solar flares reveal power-law distributions. \cite{Aschwanden_2010} and \citet{Aschwanden_2021} propose a type II Pareto distribution based on data collected over 60 years,
\begin{equation}
n_{\rm wt}(\tau)=\lambda_0(1+\lambda_0 \tau)^{-m}   ,  
\end{equation}
where $\tau$ is the time between two flares, $1/\lambda_0$ is the breaking point of the power-law.
We use the value $1/\lambda_0=0.80 \pm 0.14$ hour, as suggested by the authors. $m$ is the power-law index of the distribution. Solar observations show that $m$ ranges from $2.1$ to $2.4$. The average time between two events decreases as the power-law index increases. Since T Tauri stars are expected to be more magnetically active than the Sun, we take the upper bound of the solar observed waiting time power-law index, $m=2.4$. This gives shorter waiting times. 

From this distribution, we compute the mean waiting time between two flaring events,
\begin{equation}
    \bar{\tau}=\int_0^\infty \tau n_{\rm wt}(\tau) \rm d \tau \approx 5 ~\rm ks.
\end{equation}

\subsubsection{Luminosity Distribution}

The exhaustive study by \cite{Getman_2021a} focused on the energetics of X-ray flares from Pre-Main Sequence (PMS) stars. Their observations revealed that the X-ray peak luminosity distribution of high luminosity flares ($L>10^{32}~ \rm erg s^{-1}$) follows a power-law with an index of $2.11$. For luminosities below this threshold, the trend of the power-law cannot be deduced because the set of data is incomplete. However, this index of 2.11 corresponds to the value of the power-law index observed over a wide range of luminosities, from optical to extreme-UV (EUV) and X-ray domains, in solar and stellar flares. The range of validity of this power-law starts from solar nanoflare energies ($E_{\rm flare}=10^{24}$ erg) up to superflare energies of solar-type stars ($E_{\rm flare}=10^{35}$ erg) \citep{2019ApJ...876...58N,2021ApJ...906...72O}. This justifies our extrapolation of the power-law distribution observed by \citet{Getman_2021a} to luminosity lower than $10^{32}~ \rm erg s^{-1}$.

The normalised X-ray luminosity distribution can be written as,
\begin{equation}
    n_X (L) = \frac{a-1}{L_c} \left(\frac{L}{L_c}\right)^{-a},
    \label{eq:NormalisationLuminosityDistribution}
\end{equation}
where $a=2.11$ is the power-law index measured by \cite{Getman_2021a} and $L_c$ is the low luminosity cut-off of the distribution. $L_c$ is adjusted in order to reproduce the occurrence rate $N_{\rm MF}$ of mega flares, flares with peak X-ray luminosity $L_{X,\rm pk} > L_0=10^{32.5}$, observed by \cite{Getman_2021a}, which is $N_{\rm MF}= 1.7$ mega-flares per year per star.

In the previous paragraph we have calculated the mean waiting time between two flares $\bar{\tau}=5 ~ \rm ks $. We can estimate the total number of flares per year, $N_{\rm tot} = 6300 ~\rm{flares/year/star}$.

To find the low luminosity cut-off we start from the equation,
\begin{equation}
    N_{\rm tot}= \frac{N_{\rm MF}}{P(L_{X,\rm pk}>L_0)},
\end{equation}
where $P(L_{X,\rm pk}>L_0)$ is the probability of occurrence of a megaflare, i.e.,
\begin{equation}
    P(L_{X,\rm pk}>L_0)=\int_{L_0}^\infty n_X(L) dL= \left(\frac{L_0}{L_c}\right)^{1-a},~{a > 1}
    .
    \label{eq:LuminosityDistribution}
\end{equation}
From the two above equations, we get,
\begin{equation}
    L_c= L_0 \left(\frac{N_{\rm MF}}{N_{\rm tot}}\right)^{\frac{1}{a-1}}.
\end{equation}

Taking $N_{\rm MF}=1.7$, $N_{\rm tot}=6300$, $a=2.11$, $L_0=10^{32.5}~\rm  erg ~s^{-1}$, the low-luminosity cut-off is,
\begin{equation}
 L_c \simeq 1.9\times 10^{29}~\rm  erg ~s^{-1} . 
\end{equation}

\subsubsection{Flares luminosity profile}

We consider a simple time evolution model of flare soft X-ray luminosity $L_X(t)$ as typically observed in the Sun \citep{benz2002kinetic}. It is characterised by an exponential fast growth and gradual decay. There are three parameters in this model. The first parameter is the characteristic rise time ($\tau_r$), the second,the characteristic decay time ($\tau_d$) and, the last, the peak X-ray luminosity of the flare ($L_{X,\rm pk}$). The temporal evolution of a flare with luminosity peaking at $t=t_0$ is given by,
\begin{equation}
L_X(t) = \left\{
    \begin{array}{ll}
        L_{X,\rm pk} e^{\frac{(t-t_0)}{\tau_r}}  & \mbox{if } t<t_0 \\
        L_{X,\rm pk} e^{-\frac{(t-t_0)}{\tau_d}} & \mbox{otherwise. }
    \end{array}
\right.
\end{equation}

The peak luminosity is randomly chosen according to the distribution in the previous section. 
Figure 6 in \citet{Getman_2021a} shows no convincing correlation between the luminosity of the flares and their decay time. Thus, we derive a growth time following a Poisson distribution of the form,
\begin{equation}
    n(\tau_r)=\lambda \exp(-\lambda \tau_r),
\end{equation}
where $1/\lambda=26$ ks is the average rise time of the flares from the observed sample by \citet{Getman_2021a}. 

Figure 6 in \citet{Getman_2021a} shows a correlation between flare growth and decay times (both in ks units),
\begin{equation}
    \tau_d=5.28 \tau_r^{0.60}
    \label{eq:taudtaur}
\end{equation}

From the solar flare waiting time distributions, the peak X-ray luminosity distribution and the rise and decay time distribution of the luminosity, we can model the total X-ray luminosity emitted by a PMS. 

In solar flares, the fast rise followed by a gradual decay of the luminosity describes the soft X-ray component emitted by the thermal plasma. Conversely, the non-thermal particles, accelerated during the flare, produce hard X-rays via Bremsstrahlung and such hard X-rays are only observed during the rising phase of the soft X-ray emission \citep{2017LRSP...14....2B}. 

The time-dependent ionisation rate, represented by $\zeta(t)$, reflects the temporal profile of these non-thermal particles. Besides, the temporal profile of non-thermal particles follows the profile of hard X-rays. Consequently, we assume that the profile of the ionisation rate matches the phase of increasing soft X-ray luminosity.

\begin{equation}
\zeta(t,L_{X,\rm pk}) = \left\{
    \begin{array}{ll}
        \zeta_{\rm pk}(L_{X,\rm pk}) e^{\frac{(t-t_0)}{\tau_r}}  & \mbox{if } t<t_0 \\
        0 & \mbox{otherwise, }
    \end{array}
\right.
\label{eq:IonizationRateTemporalProfile}
\end{equation}
where $\zeta_{\rm pk}$ is the ionisation rate at the peak of the luminosity that is calculated using the method described in B23, and $t_0$ is the time at the luminosity peak. 

The waiting time, the rising time and the decaying time of the flares are of the same order of magnitude. Thus, a luminosity continuum appears. Our model produces a mean luminosity over a year of $4.0\times 10^{30}$ erg s$^{-1}$, which is close to the typical value of $ 2\times 10^{30}$ erg s$^{-1}$ of the X-ray continuum observed on PMS \citep{Flaccomio_2003,Wolk_2005}.

\subsubsection{Flare geometry Model}
Flares, characterised by sudden energy discharges within a tenuous plasma confined to a "magnetic bottle",
have been extensively explored in various works \citep{2007A&A...471..271R,2017LRSP...14....2B}. The energy dissipation in these scenarios predominantly occurs via optically thin radiation and efficient heat conduction directed towards the chromosphere of the star \citep{2014LRSP...11....4R}. Magnetic confinement plays a crucial role in determining the typical flare light curve. It yields a fast rise in soft X-ray emissions due to the rapid plasma heating, followed by a gradual, near-exponential decay resulting from combined effects of thermal conduction and radiation.

This temporal evolution follows a very general trend that the majority of observed flares satisfy. It has turned out to be a key to deduce the half-length ($L_f$) of the magnetic structure that governs the eruption \citep{2007A&A...471..271R, Serio91}. \citet{2011AJ....141..201M} proposed an improvement to the original \citet{Serio91} relationship, expressed as follows,

\begin{equation}
    L_f=\frac{\tau_d T^{1/2}}{\gamma F(\epsilon)} ,
    \label{eq:Halflength}
\end{equation}
where $\gamma \simeq 3.7~10^{-4}$ is generally valid for solar-type stars and the empirical function $F(\epsilon) = \frac{0.63}{\epsilon-0.32} +1.41 > 1$. The introduction of the function $F(\epsilon)$ includes the effect of a progressive heating in the rise phase of the flare whereas \citet{Serio91} assumed an impulsive heating with $F(\epsilon)=1$. The parameter $\epsilon$ depends on the star. It varies for a given star from one flare to the other. \citet{2011AJ....141..201M} derived the mean value of this parameter for a sample of flares in Class I, II and III young stellar objects. In this study we are interested in Class II YSO and for this class $\langle \epsilon \rangle  = 1.43$ and $\langle F(\epsilon) \rangle  \simeq 1.98$, we hence assume $F(\epsilon) = 2$ hereafter.

The half-length of the flare loop is a proxy for the position $R$ where the flux tube is in contact with the disc, thus where the EPs are released in the disc.

\begin{figure}
    \centering
    \includegraphics[width=0.7\linewidth]{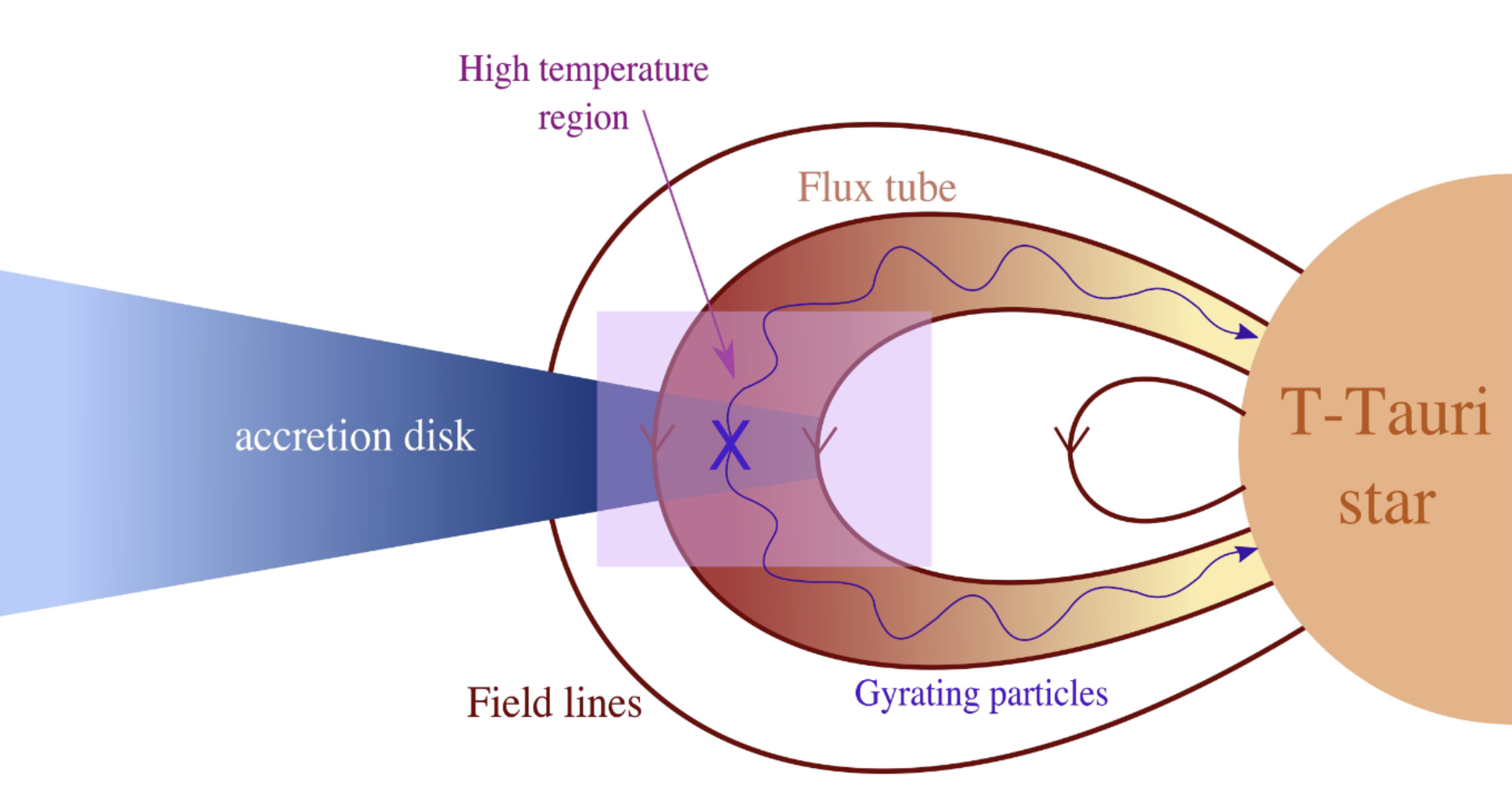}
    \caption{Sketch of a flaring flux tube reproduced from \citet{waterfall2020predicting}. Magnetic reconnection takes place at the marked X-point, in the interaction region between the star and disc magnetic field lines. Non-thermal particles move in a circular motion from this point, travelling along the field lines towards the surface of the star emitting X-rays. According to this model particles are entering the disc at $R\approx L_f$.}
    \label{fig:TTauriCoronalFlare}
\end{figure}

We will discuss two models for the estimation of the position $R$. The first model that we call the Star-Star flare (hereafter SSF) model, is supported by X-rays observations of \citet{Getman_2021b} where the two footpoints of the flare are anchored onto the central star and the flare loop-top is in the disc, see Fig \ref{fig:TTauriCoronalFlare}. \citet{waterfall2020predicting} estimated the radio emission of a flare in this configuration.

\begin{figure}
    \centering
    \includegraphics[width=0.7\linewidth]{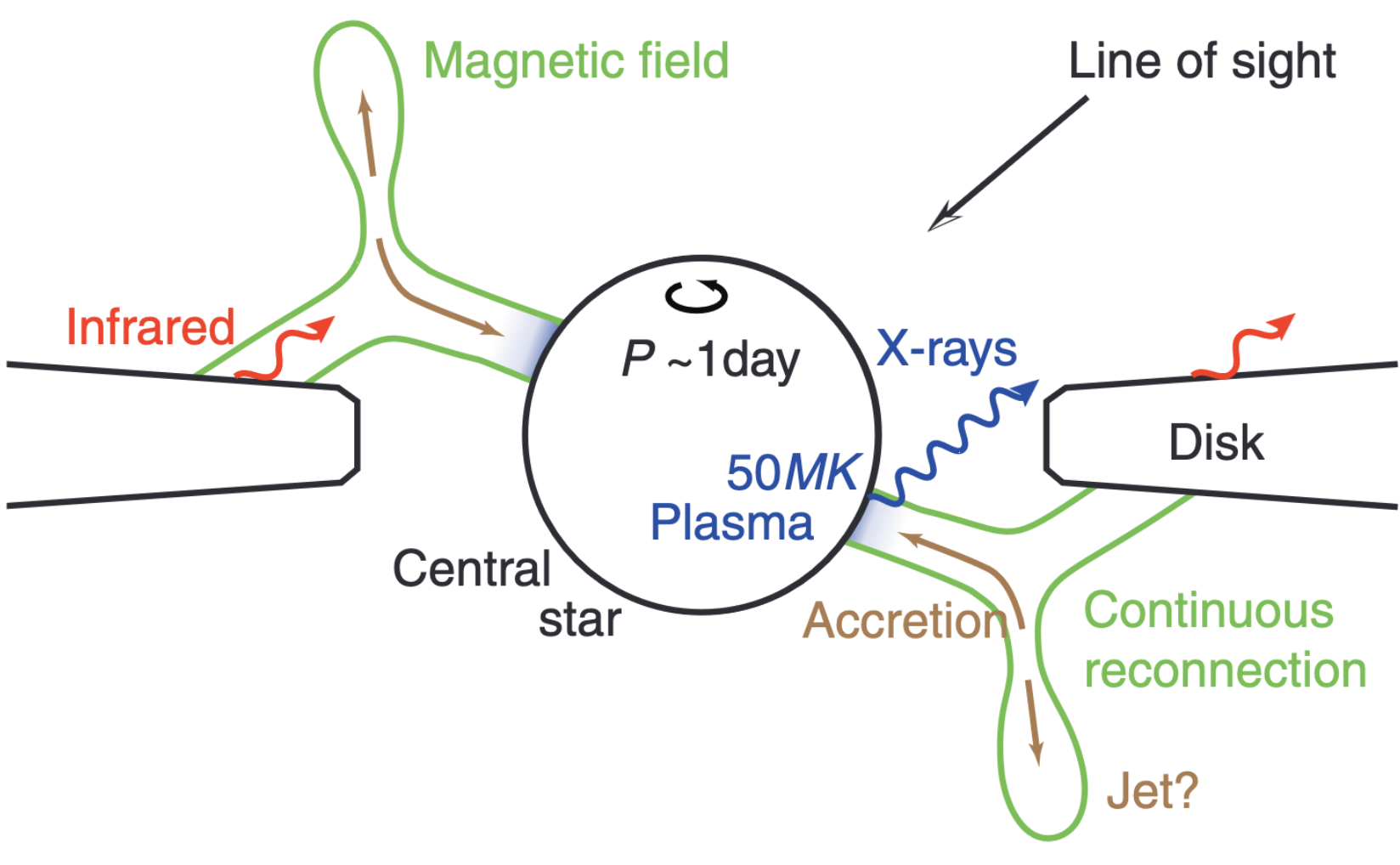}
    \caption{Possible geometry of a T Tauri flare reproduced from \citet{2012ApJ...754...32H}. Differential rotation between the star and the disc shears the stellar bipolar magnetic fields, and the magnetic fields twist and continuously reconnect. Matter, accelerated by the magnetic reconnection, collides with the stellar surface, thermalises and emits hard X-rays. According to this model particles are entering the disc at $R\approx 2 L_f$.}
    \label{fig:StarDiscFlareLoop}
\end{figure}

The second model, that we call Star-Disc flare (hereafter SDF) model, is supported by simulations \citep{2011MNRAS.415.3380O,zanni2013mhd,2019A&A...624A..50C} and by the observations of \citet{2005ApJS..160..469F} with Chandra and \citet{2012ApJ...754...32H} with XMM-Newton and Suzaku observatories. This model assumes that one flare loop is anchored to the star while the other is anchored to the disc, and the loop-top, where reconnection occurs, is above the disc, in the magneto-ejection region, see Fig \ref{fig:StarDiscFlareLoop}. 

The main impact that each of these models has on our results is the position at which the EPs produced in the reconnection region enters in the disc. Assuming the first model (SSF), with the two foot points anchored into the star, the particles are penetrating at $R\approx L_f$. On the other hand, assuming the second model (SDF) with the flare loop linking the stellar surface with the disc, particles are penetrating at $R\approx 2 L_f$. The second model injects particles farther in the disc. We show that this has a major impact on the ionisation spatial distribution in the inner disc, thus potentially to the extend of the magnetorotational instability (MRI) active region.

From Eq. \eqref{eq:TflareLX}, we can assign to each flare a half loop length given its peak X-ray luminosity $L_{X,\rm pk}$ and its decay time $\tau_d$. These values are picked from the distributions obtained in Eq.\eqref{eq:LuminosityDistribution} and Eq. \eqref{eq:taudtaur} and allow to derive a radial position $R$.

\citet{Getman_2021a} also made an estimation of $\chi$, the ratio between the flare loop half-length and its cross-sectional radius $r$. So the area of the flare cross section $A_f$ is expressed in terms of $\chi$ as, 
\begin{equation}
    A_f= \pi \chi^2 L_f^2.
    \label{eq:FlareArea}
\end{equation}
In solar and stellar flares as well as in the PMS flares observed by \citet{Getman_2021a}, the typical value of $\chi$ is 0.1. 

Each flare is given a radial location in the disc based on its luminosity, while the azimuthal location is randomly picked in an uniform distribution of angles.

\subsection{Chemical Model}
\subsubsection{ \prodimo} \label{sec:Prodimo}
To evaluate the column density explored by EPs, we use a model generated by the radiation thermochemical code \textsc{ProDiMo} \citep[PROtoplanetary DIsc MOdel\footnote{\url{https://prodimo.iwf.oeaw.ac.at} \mbox{revision: 66efbd75 2023/06/27}},][]{woitke2009radiation,kamp2010radiation,thi2011radiation,woitke2016consistent}. This model performs a wavelength-dependent radiative transfer calculation, including X-rays  \citep{Rab18}, which determines the gas and dust temperatures and the local radiation field. Chemical abundances are used to balance the heating and cooling processes, and the chemical network contains 235 different chemical species and 3143 chemical reactions \citep{kamp2017consistent,Rab17}.
Here, we use a fiducial T Tauri disc model described in \citet{Rab17}. It was also used in \citet{brunn2023ionization} (see their \mbox{Tab. 1} for the main physical parameters). Here, we want to model in a more realistic way, the thermal and ionisation structure in the innermost region of the disc ($R< 1\,\mathrm{au}$). Thus, we additionally include accretion heating following \citealt{1998ApJ...500..411D}, see also \citealt{Oberg2022} Appendix A). We use a total mass accretion rate onto the star of $\dot{M}_\mathrm{accr}=10^{-8}\,M_{\odot}\,yr^{-1}$. We added a simple treatment of the collisional ionisation to the chemical network following \citet{Desch2015}. For further details, see Appendix~\ref{app:collion}.

\subsubsection{Electron density equilibrium}
In the framework of EP propagation within the protoplanetary disc, we adopt a model based on the "test-particle" approximation. We assume that these particles exert no influence on the disc chemical structure. Nonetheless, our objective is to quantify their effect on the chemistry of the disc. To this end, we determine the electron density in regions of the disc influenced by flares.

From the differential system Eq. 7 by \citet{fromang2002ionization}, and assuming global neutrality, where the electron density equals the cation density ($n_e=n_{m^+}$) while setting aside the impact of metallicity, we can simplify the system down to a single ordinary differential equation,
\begin{equation}
\label{eq:rateeq}
\frac{ d n_e}{d t} =\zeta n_n - \beta n_e^2.   
\end{equation}

In the no-flare case, assuming steady state, the initial electron density and neutral species density, noted respectively $n_{e,0}$ and $n_{n,0}$, are computed by the \prodimo code. Upholding particle number conservation, the evolution of the neutral species density is related to the electron density and the initial conditions as,
\begin{equation}
    n_n(t)=n_{e,0}-n_e(t)+n_{n,0}.
\end{equation}
In Eq.\eqref{eq:rateeq}, the parameter $\beta=\beta(T)$ is the recombination rate from Eq. 9 of \citet{fromang2002ionization}. We note $\zeta$, the total ionisation rate, being the sum of two components, $\zeta=\zeta_{f}+\zeta_{\rm Prod}$, the stationary ionisation rate, $\zeta_{\rm Prod}$, computed by \prodimo in the scenario without flare, ensuring a stationary $n_{e,0}$. On the other hand, $\zeta_{f}$, the ionisation rate due to flares, is derived from Eq. \eqref{eq:IonizationRateTemporalProfile}. 

A global study of the EP feedback on the disc chemistry via the chemical model of \prodimo will be studied in Sect. \ref{sec:feedbackdiscchemistry}. Nevertheless, it is interesting to compute analytically the electron density via the above-outlined rudimentary chemical model. Notably, temporal dynamics remain elusive within the \prodimo framework. This simple model provides insights into the temporal trajectory of electron density modulations induced by flares, an aspect presented in Sect. \ref{sec:ionizationsingleflare}.

\subsection{Viscosity model}\label{sec:Viscosity}
As already stated, in order to develop, MRI requires a sufficiently high degree of ionisation for the gas to interact with the magnetic field. The instability can be characterised by the viscous parameter $\alpha$. 

\citet{2012MNRAS.422.2737W} pointed out that Hall diffusion could, under certain circumstances, counteract the resistive damping effects of MRI. However, numerical simulations indicate a more complex scenario \citep{2014A&A...566A..56L}. Given these uncertainties, \citet{2019A&A...632A..44T} chose to exclude the effects of Hall diffusivity in estimating the effective viscosity parameter. To compute the viscous $\alpha$ parameter, we rely on the model proposed by \citet{2019A&A...632A..44T}, which introduces an empirical non-ideal MHD MRI-driven equation for \(\alpha\) suitable for physico-chemical protoplanetary disc codes such as \prodimo. The adopted expression for \(\alpha\) is given by 
\begin{equation}
    \alpha= \left(\frac{2}{\beta_{\rm mag}}\right)^{1/2} \rm min(1,\Lambda_{\text{Ohm}}) \left[\left(\frac{50}{Am^{1.2}}\right)^2+\left(\frac{8}{Am^{0.8}}+1\right)^2\right]^{-1/2}
    ,
    \label{eq:effectiveviscosity}
\end{equation}

if \(\sqrt{\beta_{\text{mag}}} \Lambda_{\text{Ohm}} > 1\) and \(\alpha \approx 0\) otherwise. 

$Am$ represents the frequency at which neutral particles collides with ions normalised to the keplerian orbital frequency, 

\begin{equation}
Am \equiv \frac{\nu_{\rm in}}{\Omega_K} = \frac{\beta_{\rm in} n_{\rm charge}}{\Omega_K}= \frac{x_e \beta_{\rm in} n_{\rm tot}}{\Omega_K} ,
\end{equation}

where $n_{\rm charge} = x_{E}n_{\rm tot}$ is the total number density of charged species and $\beta_{\rm in} = 2 \times 10^{-9} \, \rm cm^3 s^{-1}$ is the collisional rate coefficient for ions to distribute their momentum to neutrals. Thus,
\begin{equation}
    Am\approx 1\left(\frac{x_e}{10^{-8}}\right)\left(\frac{n_n}{10^{10}~\rm cm^{-3}} \right) \left(\frac{R}{1 \rm{au} } \right)^{3/2} \ .
\end{equation}

The Ohmic diffusivity is quantified by the dimensionless Elsasser number \(\Lambda_{\text{Ohm}}\), defined as the ratio of the Lorentz force to the Coriolis force,
\[
\Lambda_{\text{Ohm}} \equiv \frac{B_z^2}{4\pi \rho \eta_O \Omega_K} \equiv \frac{v_A^2}{\eta_O \Omega_K} \equiv \left( \frac{4\pi \sigma_O}{\Omega_K} \right) \left( \frac{v_A}{c} \right)^2,
\]
where $B_z$ is the vertical component of the magnetic field, $v_A$ is the Alfvèn speed and $\rho$ is the mass density of the plasma and the Ohmic resistivity is given by \(\eta_O = \frac{c^2}{4\pi \sigma_O}\). \citet{2019A&A...632A..44T} derived an approximation for the Ohmic Elsasser number in disc regions relevant to our study where, \(\sigma_O \approx \sigma_{e,O}\) and \(x_e > 10^{-13}\):
\[
\Lambda_{\text{Ohm}} \approx 1 \left( \frac{T}{100~\rm K} \right)^{1/2} \left( \frac{10^4}{\beta_{\text{mag}}} \right) \left( \frac{R}{1 \text{ au }} \right)^{3/2} \left( \frac{x_e}{10^{-9}} \right),
\]
The magnetic term \(\beta_{\text{mag}}(R, z)\) is the ratio of the thermal pressure to the magnetic pressure, given by \citep{2019A&A...632A..44T},
\begin{equation}
    \beta_{\text{mag}}(R,Z) = \beta_{\text{mid}}  \frac{P_{\text{th}}(R,Z)}{P_{\text{th}}(R, 0)} = \beta_{\text{mid}} \frac{n(R,Z)T(R,Z)}{n(R,0) T(R,0)},
    \label{eq:defplasmabeta}
\end{equation}
where \(\beta_{\text{mid}}= \beta_{\text{mag}}(R,0)\) is the value of $\beta_{\text{mag}}$ on the midplane. \(\beta_{\text{mid}}\) is assumed to be independent of the radius so that it can be determined from the thermal structure and chemical abundances of \prodimo. 
The parameter $\beta_{\text{mid}}$ is typically in the range $10^4-10^6$. We take $\beta_{\text{mid}}=10^4$ as reference value.

\section{Results}\label{S:RESULTS}

\subsection{Single flare effects}
\subsubsection{Ionisation rates }\label{sec:ionizationsingleflare}

\begin{figure}
    \centering
    \includegraphics[width=1\linewidth]{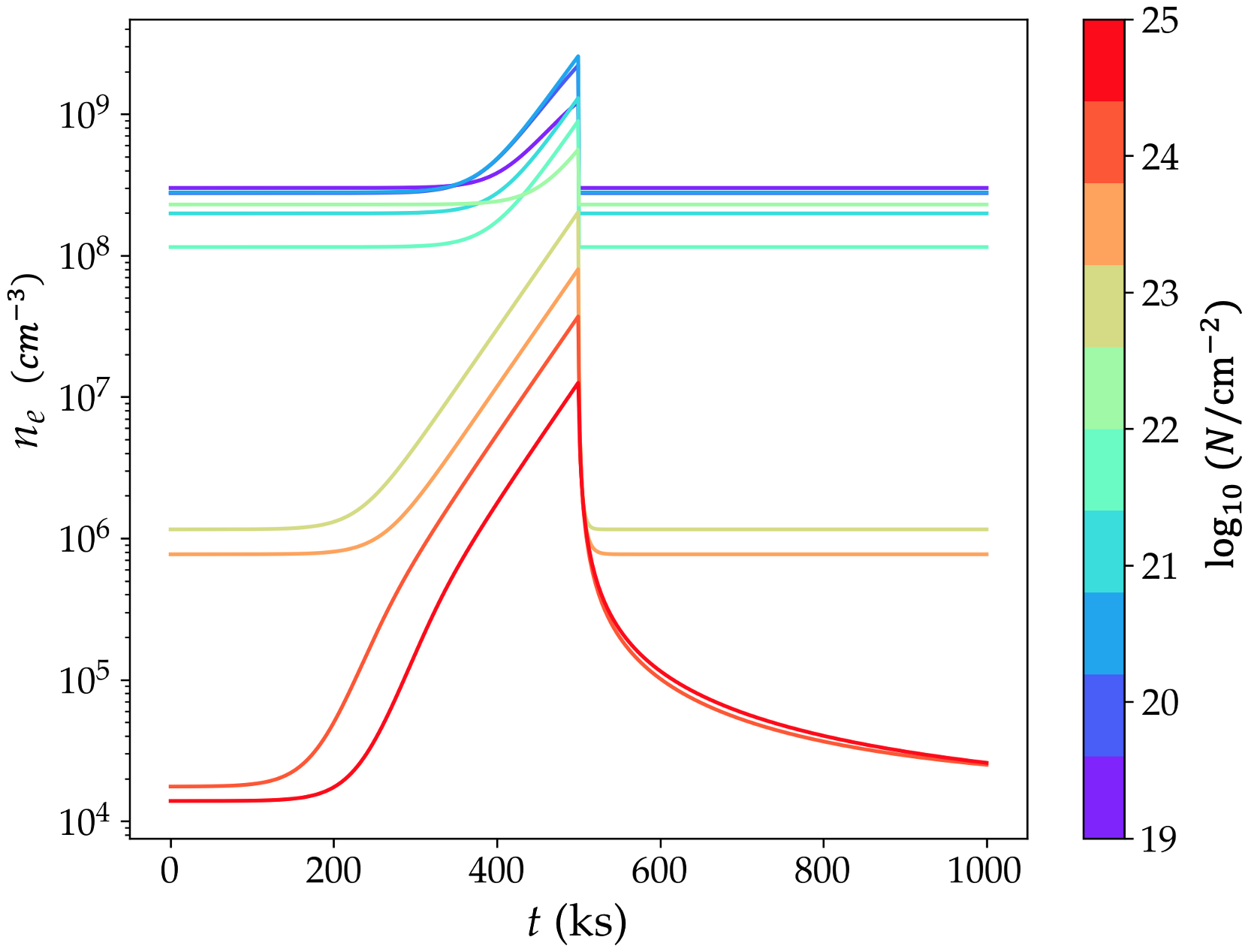}
    \caption{Plot of the electron density versus time in a range of column densities between $10^{19}$ and $10^{25}$ cm$^{-2}$. The flare temperature is 10 MK flare and the particles are penetrating into the disc at 0.1 au from the central star. The energetic particle ionisation rate follows the profile of Eq. \eqref{eq:IonizationRateTemporalProfile} with the  rising characteristic time $\tau_r=26$ ks, being the average value observed by \citet{Getman_2021a}.}
    \label{fig:Eqrateplot}
\end{figure}

We plot in Fig. \ref{fig:Eqrateplot} the solution of Eq. \eqref{eq:rateeq} for the number density of electrons. The solution is obtained by taking the time-dependent ionisation rate produced by the particles propagating along vertical field lines emitted by a single 10 MK flare at $R=0.1$ au. The electron density is computed for column densities ranging from $10^{19}~\rm cm^{-2}$ to $10^{25}~\rm cm^{-2}$. It illustrates that for column densities below $10^{24}~\rm cm^{-2}$, the system rapidly returns to a steady state. In this range of column densities, due to the high electron density, the recombination time $\tau_{\rm rec}=\left(\beta(T)n_e\right)^{-1}\sim 1$ s. Thus, the increase of the electron density only occurs during the time during which particles are injected. On the other hand, at $N>10^{24}$~cm$^{-2}$, despite the low temperature ($\approx 500 K$), the electron density is very low so the recombination times are longer, $\tau_{\rm rec}\gtrsim 10^5 s$, thus electrons remain free much longer. As a result, the effects of the flare persist for an extended period at high column density. Given that the equilibrium state discussed in this section is rather basic and solely accounts for the recombination of electrons with H$^+$, it is necessary to refine these results. This can be done especially by taking into account various other recombination processes, including those occurring on grains or metals.

In the next section, we study the effect of multiple flares using the full chemical model of \prodimo accounting for all recombination processes.

In B23, we argued that the ionisation rates obtained in the stationary state were overestimated and in order to have more realistic description of the ionisation processes, temporal and spatial averaging should be performed. We propose here a simple estimation of the spatial and temporal average of the ionisation rate due to a single flare and postpone a more precise treatment to next section. We assume the same flare as above, 10 MK at 0.1 au, with a rising time $\tau_r= 26$ ks corresponding to the average rising time observed by \citet{Getman_2021a}. We define an effective time-averaged ionisation rate $\zeta_{\rm eff}$ from the average ionisation fraction during a period of time $P$, 
\begin{equation}
    \zeta_{P,\rm eff}=\frac{1}{P}\langle x_e \rangle _ P = \frac{1}{P} \sqrt \frac{\langle\zeta\rangle _ P}{\beta n_n}  ,
\end{equation}
where $\beta$ is the recombination rate, $x_e$, the ionisation fraction and $\langle \cdot \rangle_P$ is the time averaging over a period of time $P$. The temporal averaging of the ionisation rate produced by a single flare over a period $P$ is,
\begin{equation}
    \langle \zeta_P \rangle =\frac{1}{P}\int_0^{P} \zeta(t) dt.
    \label{eq:Temporalaveraging}
\end{equation}

Given the temporal profile of Eq. \eqref{eq:IonizationRateTemporalProfile}, the average ionisation rate produced by a flare occurring at $t=t_0\ll P$ is, 
\begin{equation}
    \zeta_P=\zeta_{\rm pk}(L_{X,\rm pk}) \frac{\tau_r}{P}.
\end{equation}

\begin{figure}
    \centering
    \includegraphics[width=\linewidth]{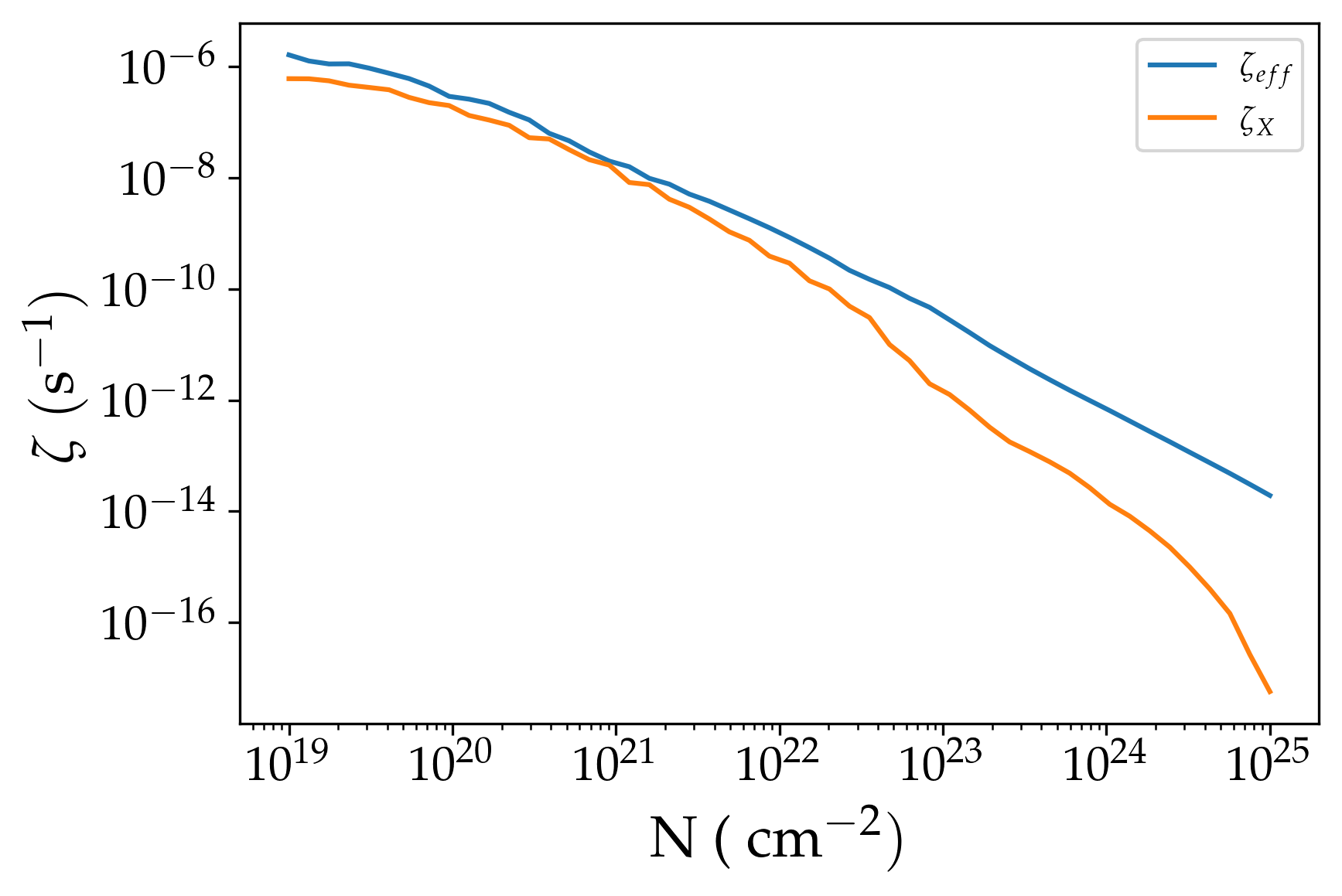}
    \caption{Ionisation rate as a function of the column density. The solid blue line shows the effective ionisation rate from a 10 MK flare averaged over a Keplerian time. In comparison, the solid orange line shows the X-ray ionisation rate of a T Tauri star as computed by \prodimo.}
    \label{fig:EffectiveIonizationRate}
\end{figure}

We compute $\zeta_{P,\rm eff}$ over a Keplerian period at 0.1 au, so $P\approx 10$ Ms. We then define the spatially averaged ionisation rate as,
\begin{equation}
    \zeta_{\rm eff}= \frac{r_f^2}{(R+r_f)^2-R^2} \zeta_{P,\rm eff},
\end{equation}
where $R$ is the radial position of flare and $r_f$ the radius of the flare. $\zeta_{\rm eff}/\zeta_{P,\rm eff} $ is the ratio between the area illuminated by particles and the disc ring of width $2 r_f$ at radius $R$. Here we take $R=0.1$ au and $r_f/R=0.1$. This effective ionisation rate is illustrative for comparison with other sources of ionisation in the inner disc. The effective ionisation rate is plotted in Fig. \ref{fig:EffectiveIonizationRate} and compared to the X-ray ionisation rate computed by \prodimo.
It was anticipated that calculating spatial and temporal averaged ionisation rates would result in lower values than the ionisation rate estimated at a flare maximum in B23. This hypothesis is validated here. Considering that observations span over extended periods of time, the averaged ionisation rates calculated in this study offer a better basis for comparison with observational data than the ionisation rate obtained in B23. Furthermore, these averaged ionisation rates also provide more relevant values to be used in numerical models of stationary chemistry in discs such as \prodimo.
As seen in Fig. \ref{fig:EffectiveIonizationRate}, the effective ionisation rate produced by flares at low column densities, \( N\lesssim 10^{22}~\rm cm^{-2}\), is similar to the ionisation rate produced by X-rays. However, at high column densities, \( N\gtrsim 10^{22} \) cm\(^{-2}\), the effective ionisation rate produced by flares becomes one to two orders of magnitude higher than that produced by X-rays. The increase in ionisation rate by EPs compared to X-rays at high column densities is an expected result as has been demonstrated in \citet{Rab17, Padovani18, 2019ApJ...883..121O}.

\subsubsection{Non-thermal pressure}
\begin{figure}
    \centering\includegraphics[width=\linewidth]{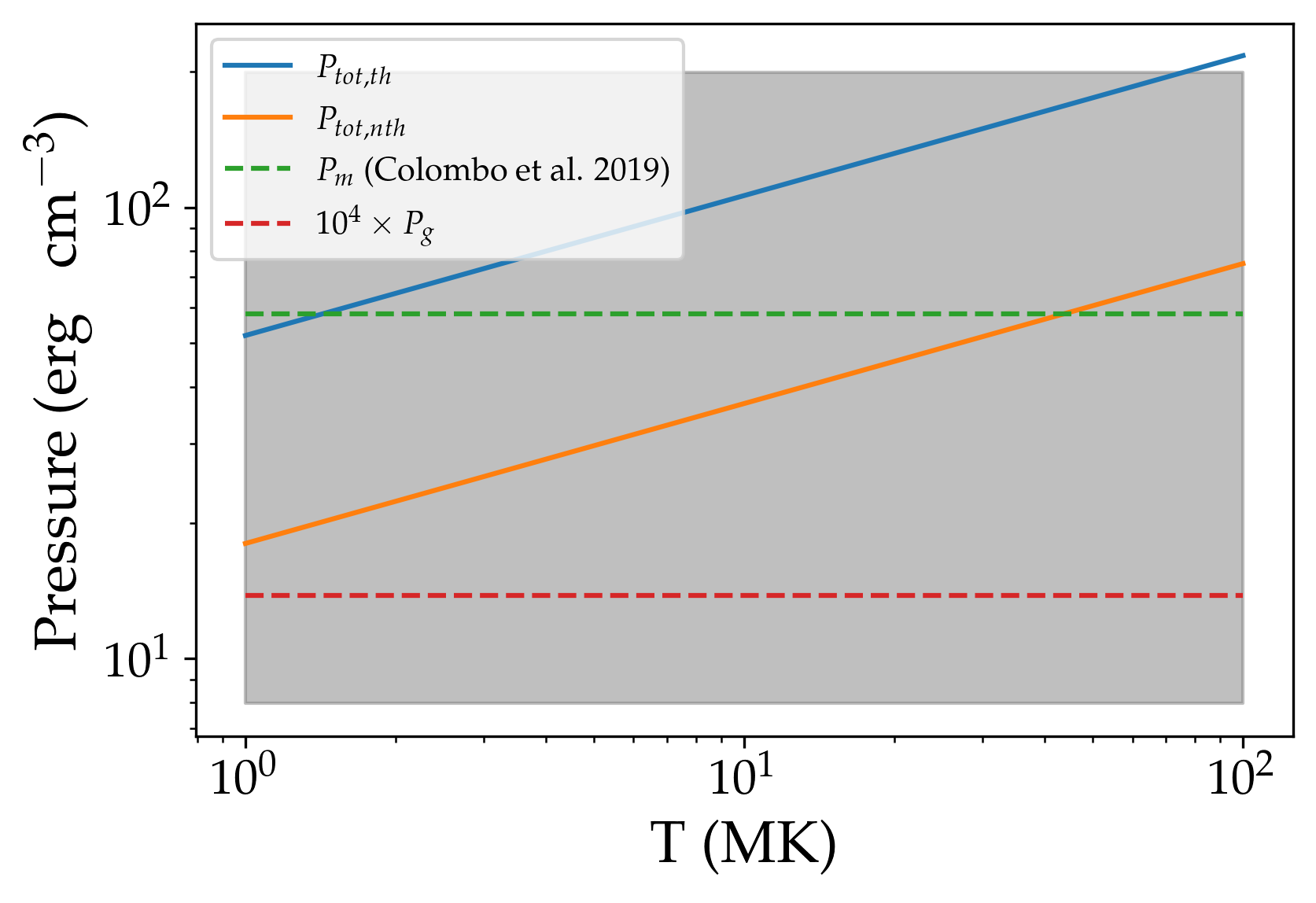}
    \caption{Different components of the pressure in the disc as a function of the flare temperature in the range 1$-$100~MK: pressure of thermal particle distribution produced by flares (solid blue line); pressure of the suprathermal component of the particle distribution (solid orange line); magnetic pressure resulting from a 27 G magnetic field, as used in the simulation of the thermal effects of flares by \citet{2019A&A...624A..50C} (dashed green line); thermal pressure at the disc surface, with \(T=10^3\) K and \(n=10^{10}\) cm\(^{-3}\) (dashed red line). The grey shaded area shows the range of magnetic pressure exerted by a magnetic field with intensities between 10 and 50 G.}
    \label{fig:nonthermalpressure}
\end{figure}

In addition to altering the disc chemical state, non-thermal particles also influence its dynamics. To gauge this impact, we compare the pressure exerted by the non-thermal particle distribution to the magnetic and gas pressures. The non-thermal pressure, \( P_{i,\text{nt}} \), exerted by a distribution of of electrons and protons ($i=e,p$), can be expressed as,
\begin{equation}
    P_{i,\text{nt}} = \frac{1}{3} \int_{E_c}^\infty p(E) v(E) F_{i,\text{nt}}(E) \, dE,
\end{equation}
where \( E \) is the kinetic energy, $E_c$ is the injection energy of the non-thermal particles, and \( F_{i,\text{nt}}(E) \) is the energy distribution of these non-thermal particles \( i \),

\begin{equation}
    F_{i,\text{nt}}(E)= N_{i,\text{nt}} \left(\frac{E}{E_c}\right)^{-\delta}.
\end{equation}

We assume equipartition of energy between non-thermal protons and electrons, i.e \( F_{e,\text{nt}} = F_{p,\text{nt}} \).

We consider both species with the same injection energy $E_c$. Thus, $P_{e,\text{nt}}=P_{p,\text{nt}}=P_{\text{nt}}$, with 
\begin{equation}
    P_{\text{nt}}\simeq \frac{2}{3} \frac{\delta-1}{\delta-2} E_c n_{\rm nt},
\end{equation}
where $n_{\rm nt}$ is the non-thermal particle density (Eq. \ref{eq:nonthermaldensityfiducial}) and $\delta \ne 2$. We assume that the injection energy is proportional to the thermal energy, $E_c=\theta E_{\rm th}$, where $E_{\rm th}=\frac{3}{2} k_B T$, $k_B$ is the Boltzmann constant and $T$ the flare temperature. Then, we express the pressure of the non-thermal distribution in terms of the pressure of the thermal distribution, 
\begin{equation}
    P_{\text{nt}}=\frac{3\theta^2}{\delta-2}\sqrt{\frac{3\theta}{2\pi}}\exp(-\frac{3\theta}{2}) P_{\text{th}}.
\end{equation}
The thermal pressure can be computed as function of the flare temperature from $P_{\rm th}= n_{\rm th}(T) k_B T$, where $n_{\rm th}(T)$ is the plasma thermal density constrained from X-ray observation, Eq. \eqref{eq:thermaldensityfiducial}.  

Assuming the fiducial values of B23, $\theta=3$ and $\delta=3$,
\begin{equation}
P_{\text{th}}= 26 \left(\frac{T}{1~\rm MK}\right)^{0.31} \rm ~ erg ~ cm^{-3} \quad \text{and} \quad  P_{\text{nt}}= 9 \left(\frac{T}{1~\rm MK}\right)^{0.31} \rm ~ erg ~ cm^{-3}. 
\label{eq:thermalnonthermalpressure}
\end{equation}

The total pressure of non-thermal particles accounting for both electrons and protons is $P_{\text{tot,nt}} \approx 2 P_{\text{nt}}$ and the total thermal pressure is $P_{\text{tot,th}} \approx 2 P_{\text{th}}$.

In Fig. \ref{fig:nonthermalpressure}, we plot the total thermal and non-thermal particle pressure alongside the gas and the magnetic pressures. Above the inner disc surface, where a flare is occurring, the magnetic field intensity should be in the range  \(10-50\) G \citep{2011MNRAS.415.3380O,2019A&A...624A..50C}. The graph indicates that the pressure of thermal and supra-thermal particles generated by flares is typically comparable to the magnetic pressure at the inner disc surface. Meanwhile, the particle pressure greatly exceeds the disc gas pressure. Therefore, these particles are expected to influence the magnetic and gas structures of the discs. Earlier simulations examining the effect of pressure from flares focused on the pressure contribution from the thermal particles \citep{2011MNRAS.415.3380O,2019A&A...624A..50C}. They assumed a magnetic intensity of 27 G in the inner disc, resulting in the magnetic pressure represented by the green dashed line in Fig. \ref{fig:nonthermalpressure}. \citet{2019A&A...624A..50C} demonstrated that the pressure from the thermal component could greatly alter the thermal structure of the disc. It can even initiate overpressure waves that cross the disc, forming accretion funnels on the opposite side of the disc. 

From Eq. \eqref{eq:thermalnonthermalpressure}, we have shown that the non-thermal component accounts for over a third of the thermal pressure. Thus, it represents a non-negligible portion of the particle pressure exerted by particles from a flare. It appears essential for future flare simulations in T Tauri stars to consider the effect of the supra-thermal component.

\subsection{Ionisation by multiple flares}
A single flare appears to cause a temporary change in the disc chemistry, but the system comes back to its original chemical state over time. 
\citet{2022ApJ...928...46W} observe that the collective effects the X-rays produced by flares can push some chemical species into a "new pseudo-steady state" if their formation or destruction timescales are longer than the intervals between flares. For instance, in their study, if the creation of a particular species is boosted by a byproduct of a flare and its decay time exceeds the time between strong flares, the system will not come back to the original state before the next flare occurs. The specific timescales for these changes vary depending on the molecule in question and the prevailing local conditions. \citet{2006A&A...455..731I} also argued that for flares, the chemistry of ions in the disc should be modelled using time-averaged ionisation rates. 

So, in order to have a more complete understanding of the effects of EPs produced by multiple flares, we compute the spatial and temporal averaged ionisation rate. We use a sample of flares that are randomly drawn using the method described in Sect. \ref{seq:FlareModeling}, over a time period $P$. 

First, we need to deal with the temporal averaging of the ionisation rate produced by a single flare over a period $P$, as in Eq. \eqref{eq:Temporalaveraging}.

Given the temporal profile of Eq. \eqref{eq:IonizationRateTemporalProfile}, the average ionisation rate produced by a flare occurring at $t=t_0$ is, 
\begin{equation}
    \zeta_P=\zeta_{\rm pk}(L_{X,\rm pk}) \frac{\tau_r}{P} \left(1-e^{-\frac{P-t_0}{\tau_r}}\right)
\end{equation}

We now calculate the spatial average of the ionisation rate produced by the whole sample of flares. The spatial average is computed in disc rings with a radius $R$ and width $\Delta R$. The area of such a ring is $A_R= \pi \left(R^2-(R-\Delta R)^2\right)$. 

The flux tube of each flare $i$ from the sample intersects the disc ring with an area $A_i$, where $A_i$ is smaller than $A_f$. The flare cross section with the whole disc is given by Eq. \eqref{eq:FlareArea}. 
Flares produce a time averaged ionisation rate $\zeta_{P,i}(N)$ in a volume of section $A_i$. 

The spatial average ionisation rate $\zeta_{P,R}$ in a disc ring with a large radius $R$ is given by,
\begin{equation}
\zeta_{P,R}(N)=\frac{1}{A_R}\sum_i {A_i \zeta_{P,i}(N)}.
\end{equation}

\begin{figure}
\begin{subfigure}{.5\textwidth}
  \centering
    \includegraphics[width=1\linewidth]{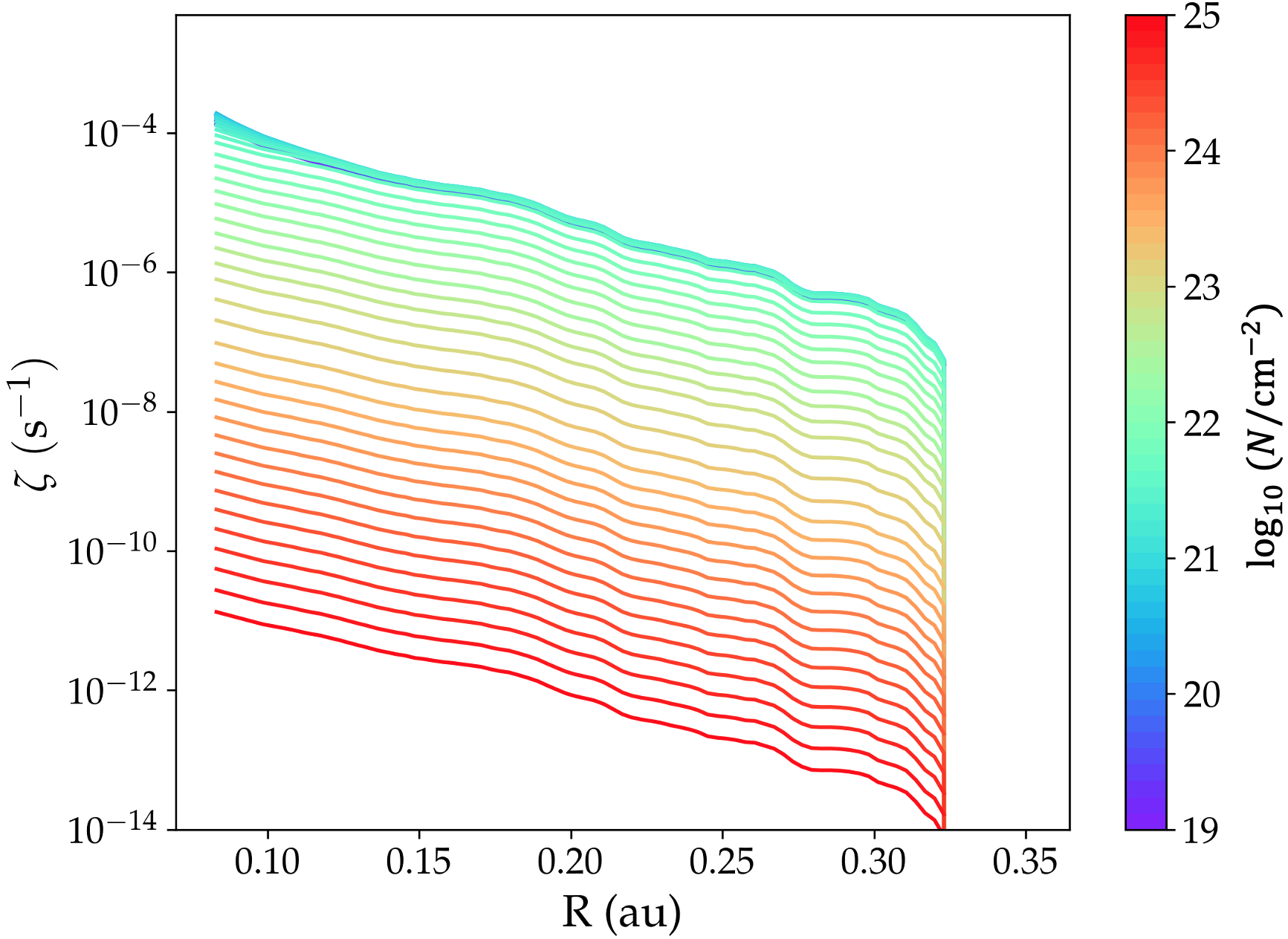}
    \caption{SSF model ionisation rates}
    \label{fig:CoronalFlareIonizationRates}
\end{subfigure}
\begin{subfigure}{.5\textwidth}
  \centering
    \includegraphics[width=1\linewidth]{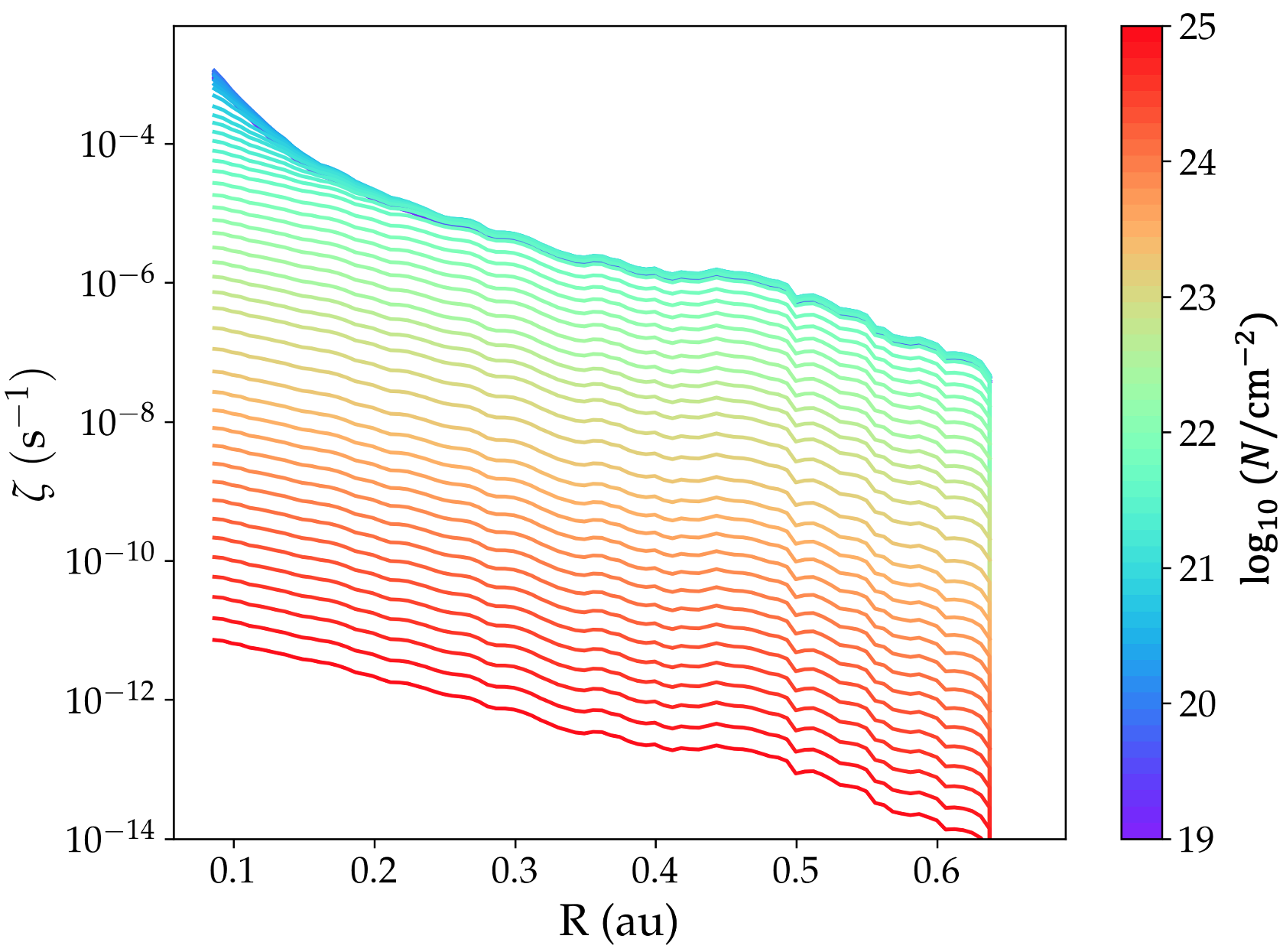}
    \caption{SDF model ionisation rates}
    \label{StarDiscFlareIonizationRates}
\end{subfigure} 
\caption{Time-averaged ionisation rate in the range $10^{19}-10^{25} ~ \rm cm^{-2}$ over a period of 1000 yr. Panel (a) shows the results obtained from the SSF model. Panel (b) shows the results from the SDF model.}
\label{fig:TimeAveragedIonizationRate}
\end{figure}

We have calculated the average ionisation rate by varying the duration $P$. On shorter time scales, the likelihood of encountering a powerful flare capable of emitting particles at large distances from the central star is relatively low. Consequently, the primary variation introduced by changing the period is the radial extent of the region impacted by flares. 
The spatial average ionisation rate plotted in Fig. \ref{fig:CoronalFlareIonizationRates} is obtained by taking $\Delta R \ll R $ and averaging over 1000 years.

From the numerical results we propose a parametrisation of the average ionisation rate due to the EPs produced by flares of the form,
\begin{equation}
    \zeta(R,N)= \zeta_0 \left(1+\frac{N}{N_0}\right)^{-a} e^{-b (R-R_0)}.
    \label{eq:ionizationrateparametrization}
\end{equation}
The parameters for the SSF and SDF models are presented in Tab. \ref{tab:ionizationrateparmeters}. This expression provides a fit matching the numerical results with an average relative deviation of less than 10\% for column densities ranging from $10^{19}$ to $10^{25}$ cm$^{-2}$, for radii between $0.1-0.3$ au and $0.1-0.6$ au for SSF and for SDF, respectively.
\begin{table}
    \centering
    \begin{tabular}{c|c|c|c|c|c}
        \hline \hline
         Parameters & $\zeta_0$ [s$^{-1}$] & $N_0$ [cm$^{-2}$] & $a$ & $b$ [au$^{-1}$] & $R_0$  [au] \\
         \hline
        SSF & $1.5\times 10^{-4}$ & $1.2\times 10^{22}$ &2.35 &26.5 &0.075\\
        SDF & $4.6\times10^{-5}$ & $7.9\times 10^{21}$&2.28&11.3&0.15\\
       \hline
    \end{tabular}
    \caption{Parameters of the ionisation rate expression Eq. \eqref{eq:ionizationrateparametrization}}
    \label{tab:ionizationrateparmeters}
\end{table}

\subsection{Feedback on the disc}\label{sec:feedbackdisc}

\subsubsection{Effects on the disc chemistry}\label{sec:feedbackdiscchemistry}
\begin{figure*}
\begin{subfigure}{.3\textwidth}
  \centering
  \includegraphics[width=1\linewidth]{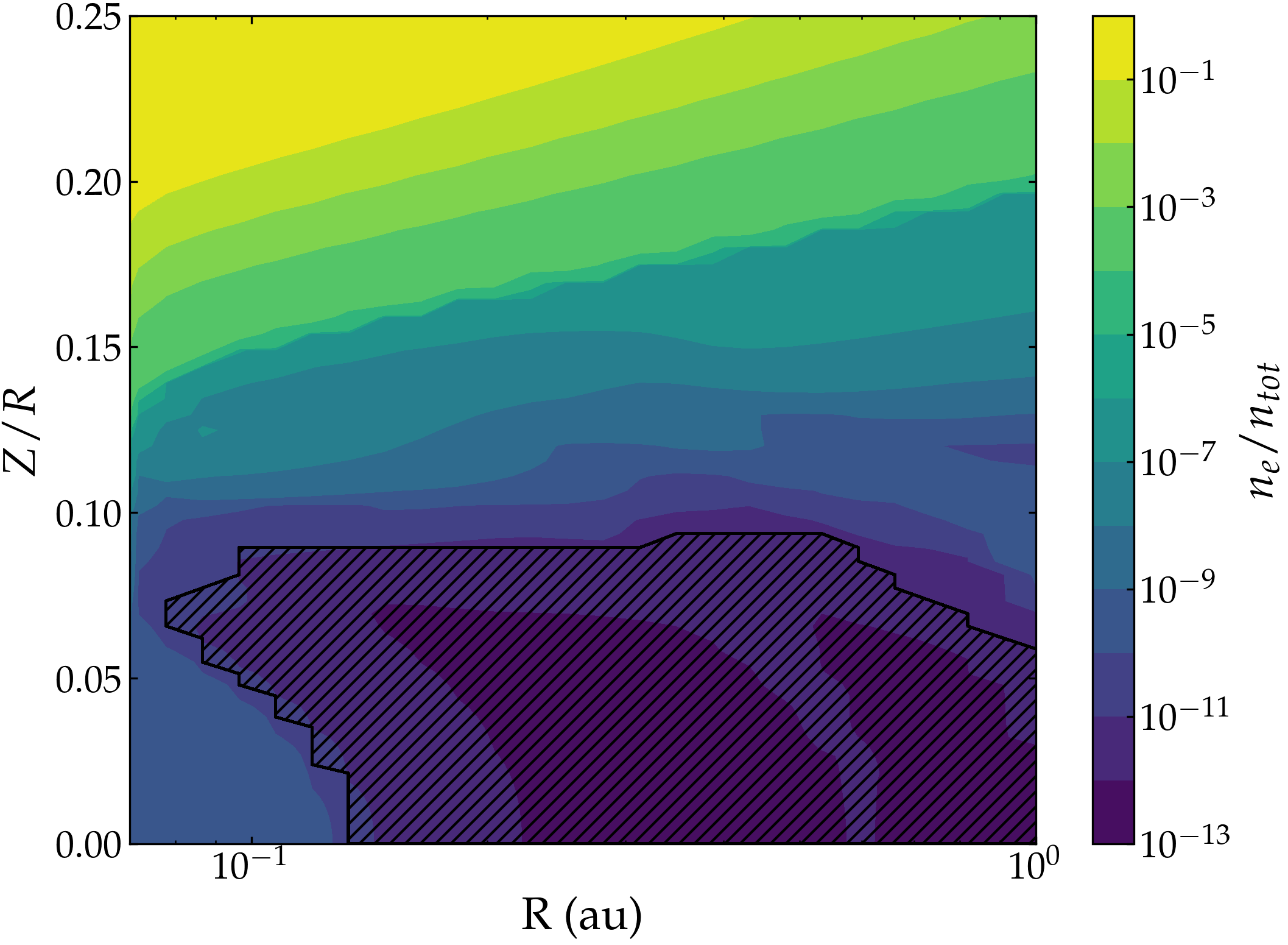}
  \caption{No Flare}
  \label{fig:ProdimoIonizationFractionNoFlare}
\end{subfigure}%
\begin{subfigure}{.3\textwidth}
  \centering
  \includegraphics[width=1\linewidth]{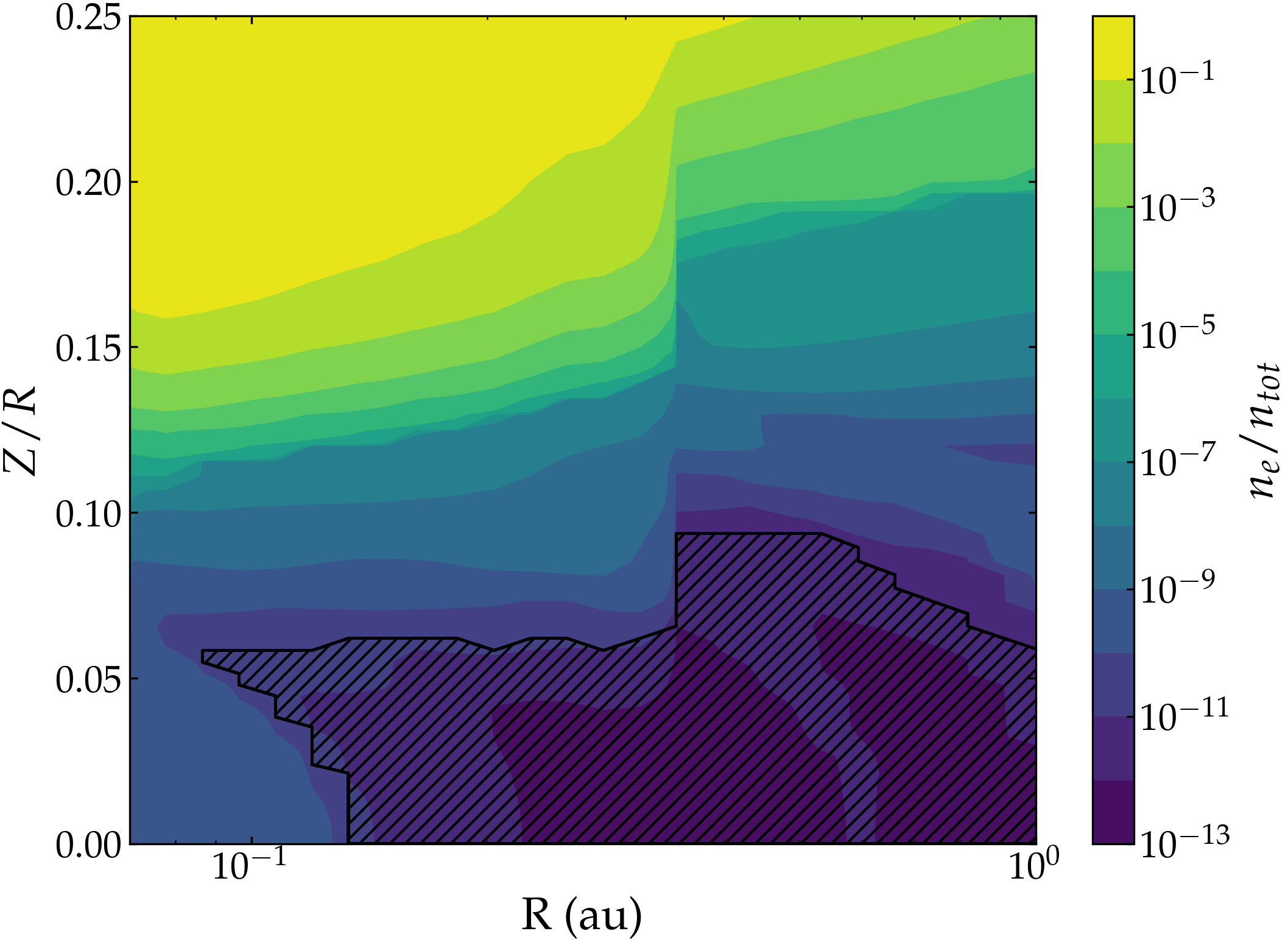}
  \caption{SSF}
  \label{fig:ProdimoIonizationFractionCoronal}
\end{subfigure}
\begin{subfigure}{.3\textwidth}
  \centering
  \includegraphics[width=1\linewidth]{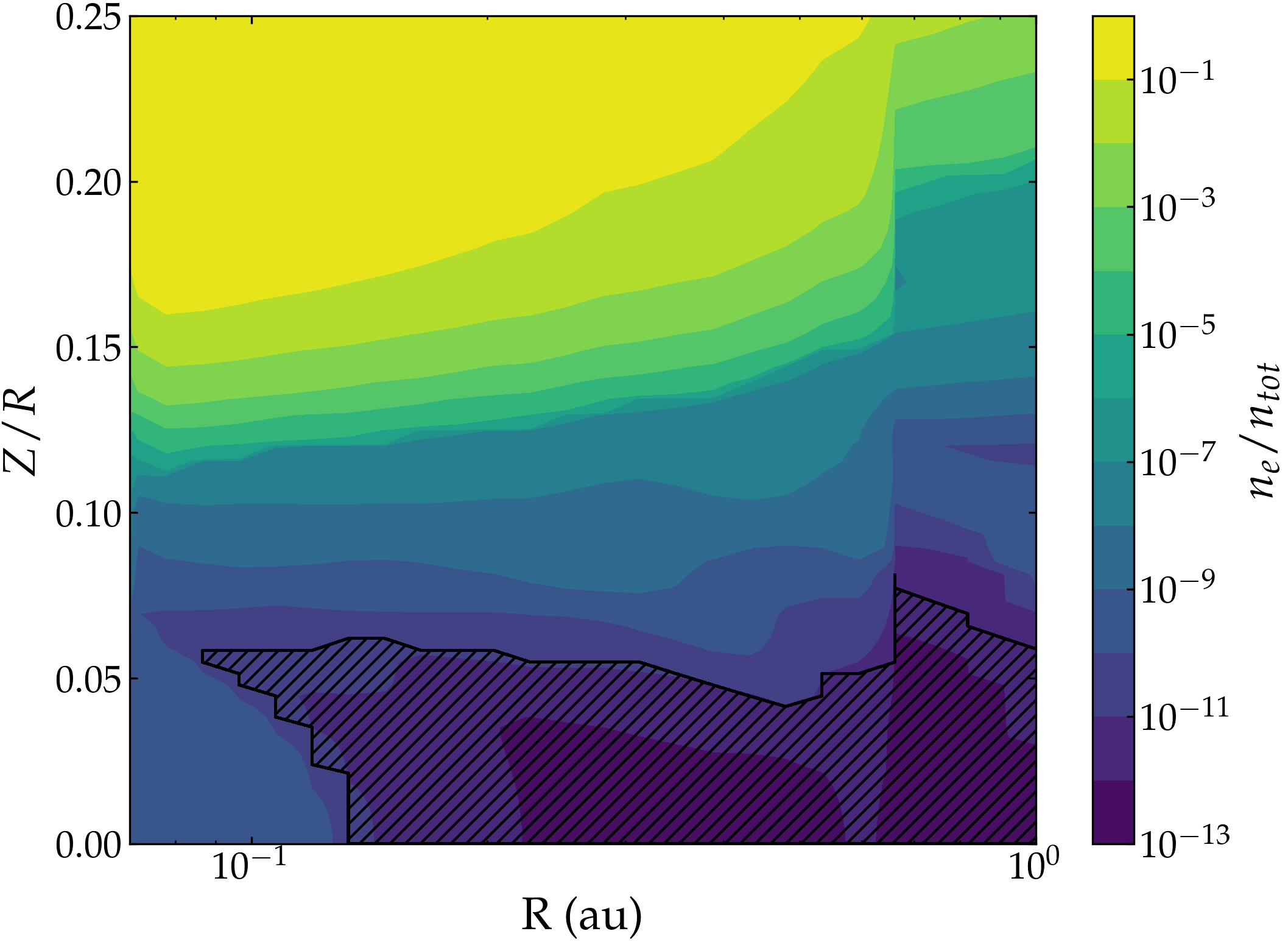}
  \caption{SDF}
  \label{fig:ProdimoIonizationFractionStarDisc}
\end{subfigure}%

\caption{Expected distribution of the ionisation fraction as a function of the disc radius, $R$, and the normalised height, $Z/R$, in the case of no flare (panel a), SSF (panel b), and SDF (panel c). The hatched areas show the MRI inactive regions. }
\label{fig:ProdimoIonizationFraction}
\end{figure*}

\begin{figure*}
\begin{subfigure}{.3\textwidth}
  \centering
  \includegraphics[width=1\linewidth]{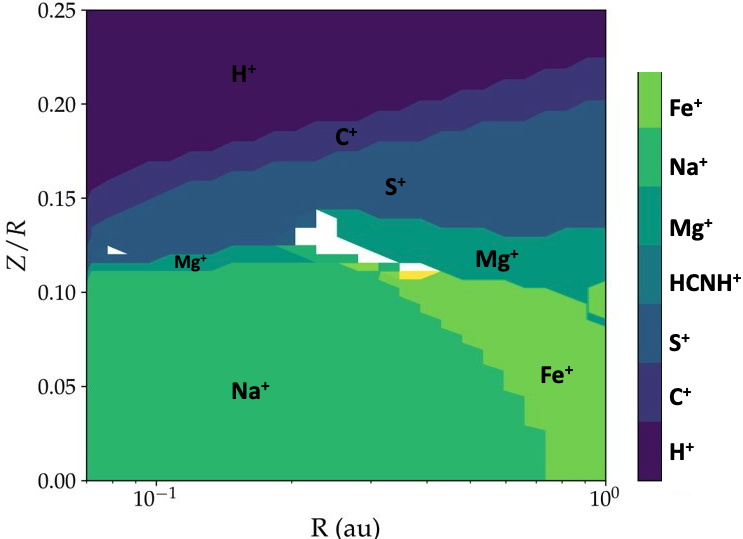}
  \caption{No Flare}
  \label{fig:CationDistributionNoFlare}
\end{subfigure}%
\begin{subfigure}{.3\textwidth}
  \centering
  \includegraphics[width=1\linewidth]{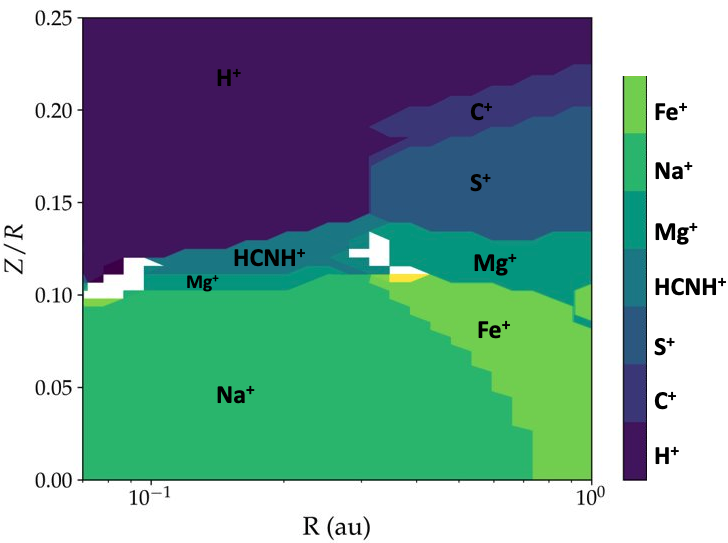}
  \caption{SSF}
  \label{fig:CationDistributionCoronal}
\end{subfigure}
\begin{subfigure}{.3\textwidth}
  \centering
  \includegraphics[width=1\linewidth]{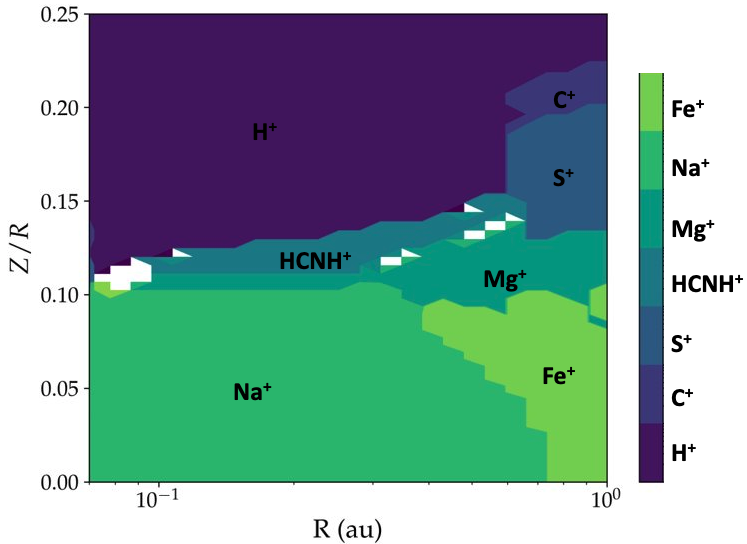}
  \caption{SDF}
  \label{fig:CationDistributionStarDisc}
\end{subfigure}%

\caption{Expected distribution of the most abundant cation as a function of the disc radius, $R$, and the normalised height, $Z/R$, in the case of no flare (panel a), SSF (panel b), and SDF flare (panel c). The white areas are regions where other cations are dominant like H$_3$O$^+$,NH$_4^+$ or SiOH$^+$.}
\label{fig:CationDistribution}
\end{figure*}

In this section, we explore how the ionisation rate, resulting from flaring activities, influences the chemical composition of the disc. This impact is quantified using ProDiMo, which takes as inputs the spatial distribution of the ionisation rate generated by both SSF and SDF represented in Fig. \ref{fig:TimeAveragedIonizationRate}. ProDiMo then recalculates the entire disc chemistry, accounting for this new ionisation source. Our model computes ionisation rate for column densities $N< 10^{25}$ cm$^{-2}$. At higher column density the ionisation rate is extrapolated using linear interpolation in log space. The revised disc structures are displayed in Fig. \ref{fig:ProdimoIonizationFraction} where we specifically examine the ionisation fraction. In Fig. \ref{fig:CationDistribution}, we examine the distribution of the most abundant cation in the inner disc. On these figures, we plot the effects of three flare models on disc structures, with the horizontal axis indicating the disc radius \( R \) in astronomical units (au) and the vertical axis representing \( Z/R \). We show the reference disc structure without flare contributions and those incorporating effects from SSF and SDF. 

In addition to the ionisation fraction, in Fig. \ref{fig:ProdimoIonizationFraction}, the solid black line shows the delimitation of the MRI inactive region, where $\alpha=0$ (i.e. where $\sqrt{\beta_{\rm mag}}\Lambda_{\rm Ohm}<1$). We assume $\beta_{\rm mid}=10^4$ (see Eq. \eqref{eq:defplasmabeta}). The method to determine the MRI active region is explained in Sec. \ref{sec:Viscosity}, and further discussed in Sec. \ref{sec:feedbackMRI}.

The overall ionisation fraction within the disc is determined by the equilibrium between ionisation sources, namely UV, X-rays, cosmic rays, and EPs from flares, and recombination as well as charge exchange reactions. The principal charge carriers in this system are free electrons for negative charges and gas-phase cations for positive charges.

Panels (b) and (c) in Fig. \ref{fig:ProdimoIonizationFraction} reveal that, in disc structures influenced by flaring activity, the ionisation fraction is higher by more than one order of magnitude compared to structures that do not account for flares (panel (a)) at heights where the Z/R ratio is between 0.05 and 0.2.

Due to the longer loops of SDF compared to SSF, the impact of the former on the ionisation fraction extends farther radially, see  Fig. \ref{fig:ProdimoIonizationFraction}. Specifically, the region in which the ionisation fraction increases by more than one order of magnitude compared to the structure without flares, extends up to $\sim 0.6$ au in structures that account for SDF. Conversely, it is limited to $\sim 0.3$ au in structures accounting for SSF. The sharp profile observed in the ionisation fraction distribution at 0.3 au and 0.6 au in Fig. \ref{fig:ProdimoIonizationFraction} (b) and (c), respectively, can be attributed to the maximum distances from the star that particles produced by the flares can reach. 

Panel (a) in Fig. \ref{fig:CationDistribution} shows the distribution of the most abundant ion species in the inner disc without flares. At the disc surface, the ionising flux arising from the Far Ultraviolet (FUV) photons is predominant, and the gas layer essentially acts as a photo-dissociation region (PDR). In this region, atomic ions like H$^+$, C$^+$, S$^+$, and Mg$^+$ are the most abundant cations.

While in the UV and X-ray shielded inner disc, collisions due to high temperature ionise atoms having low ionisation potential like the alkali, producing a region where Na$^+$ is the dominant cation.

However, injection of non-thermal particles have additional effects on the chemistry. In the region undergoing the increased ionisation rate, at $R < R_i$, where $R_i=0.3$ au  and $R_i=0.6$ au for SSF and SDF, respectively, the chemical structure is similarly modified for both flare models. The disc surface layer is no more a standard PDR, dominated by C$^+$ and  S$^+$, but the most abundant ion is H$^+$. However, a surprising feature appears. There is a HCNH$^+$ dominated layer, that is absent in the model without flare.

An important aspect for the chemistry in context of variable flares are the timescales attached to chemical reactions. In case of the EP-induced ionisation the picture is twofold. The ionisation due to the particles does not affect only the upper UV-dominated layers, but also deeper regions of the disc (towards the disc midplane). These two regions can roughly be separated by the location of the $A_\mathrm{V}=1$ layer (see e.g. Fig.~\ref{fig:disciongas}). As a consequence the region affected by the particles also spans a huge range in chemical timescales. In the UV-dominated layers those timescales a very short (about 1 day), whereas towards the midplane they span from $\approx 10^3$ to $\gtrsim 10^6$ years \citep[see ][Fig. 13, and Eq. 117 for the definition of the chemical timescale in our model]{woitke2009radiation}.

In our model the HCNH$^+$- dominated layer crossed the $A_\mathrm{V}=1$ layer and hence an interpretation of the chemistry in this region is complex. One of the consequences of the increase in the abundance of HCNH$^+$ is an enhancement of HCN abundances through HCNH$^+ + e \longrightarrow$ HCN + H \citep[e.g.][]{2021A&A...647A.118L}. This process is most important in deeper layers (below the $A_\mathrm{V}=1$ limit) of the disc with longer chemical timescales and is a consequence of the ion-molecular formation pathway as described in \citet[][see e.g. their Fig. 5]{Walsh2015} which also effects other molecules such as CN. \citet{Woitke2023} pointed out that X-ray ionisation plays an important role for the formation of HCN but also C$_2$H$_2$ in UV shielded regions. However, compared to X-rays, an irradiation by EPs produce higher ionisation rates deeper in the disc and hence those formation pathways become even more important.

\begin{figure}
    \centering
    \includegraphics[width=1\linewidth]{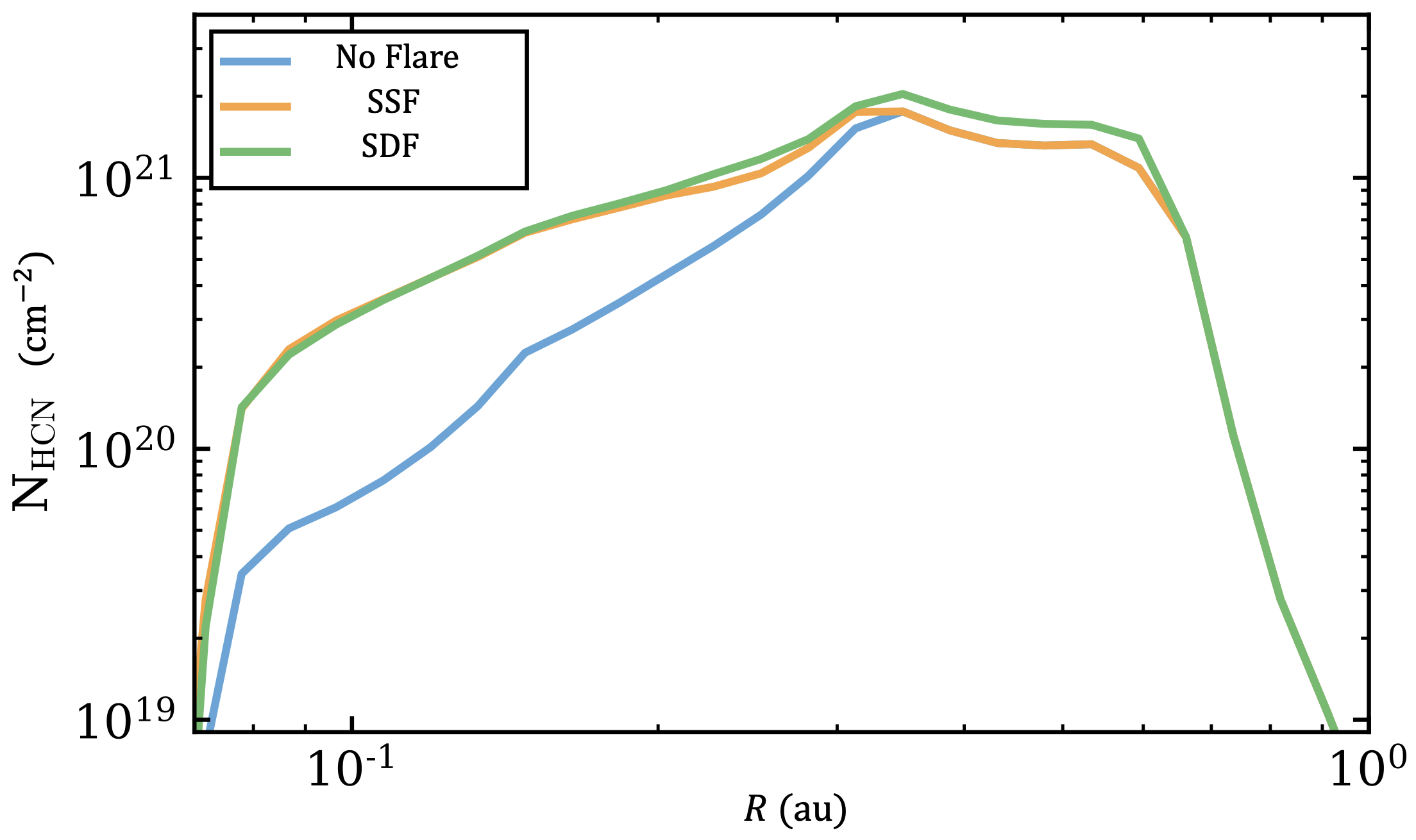}
    \caption{Vertically integrated column densities for HCN in the inner one au, for models without flares, with SDF and SSF.}
    \label{fig:coldHCN}
\end{figure}

In Fig.~\ref{fig:coldHCN} we show the HCN column density for models with and without flares. This result is given by the \prodimo steady-state chemistry model assuming averaged particle ionisation rates from the flares. As the chemical timescales are long in the deeper layers of the disc (which dominate the column density), we expect that this result does not depend strongly on the flare frequency, but more importantly on the period of time in which the disc suffers from enhanced particle irradiation due to flares. However, a more detailed analysis of the complex chemistry and the time-dependent behaviour of the disc is required to make more firm conclusions and will be treated in a forthcoming work.

Nevertheless, the enhancement of HCN is a promising outcome of our study and interesting in two ways. Firstly, the increase in HCN abundance due to flare suggests that this species may become a candidate tracer for comparing our model with observational data. Secondly, on the theoretical side, further hydrolysis of HCN may produce amino-acids. Indeed, several chemical mechanisms have been suggested for the abiotic formation of prebiotic molecules. For instance, amino acids are commonly believed to originate from such processes. An aldehyde interacts with HCN and NH$_3$, resulting either in an aminonitrile, which is then hydrolysed to produce the corresponding amino acid or in a reaction between an aldehyde, HCN, NH$_3$ and CO$_2$ that leads to the production of amino acids through hydrolysis \citep{taillades1998n}. Our model, accounting for the ionisation by flares, produces a strong increase in the HCN abundance in region at distance $R= 0.1 - 1$ au from the central star. It may be of interest for the synthesis of the "building blocks of life" on primitive Earth-like systems.

\subsubsection{Effects on MRI and Accretion Rates}\label{sec:feedbackMRI}
\begin{figure*}
\begin{subfigure}{.3\textwidth}
  \centering
  \includegraphics[width=1\linewidth]{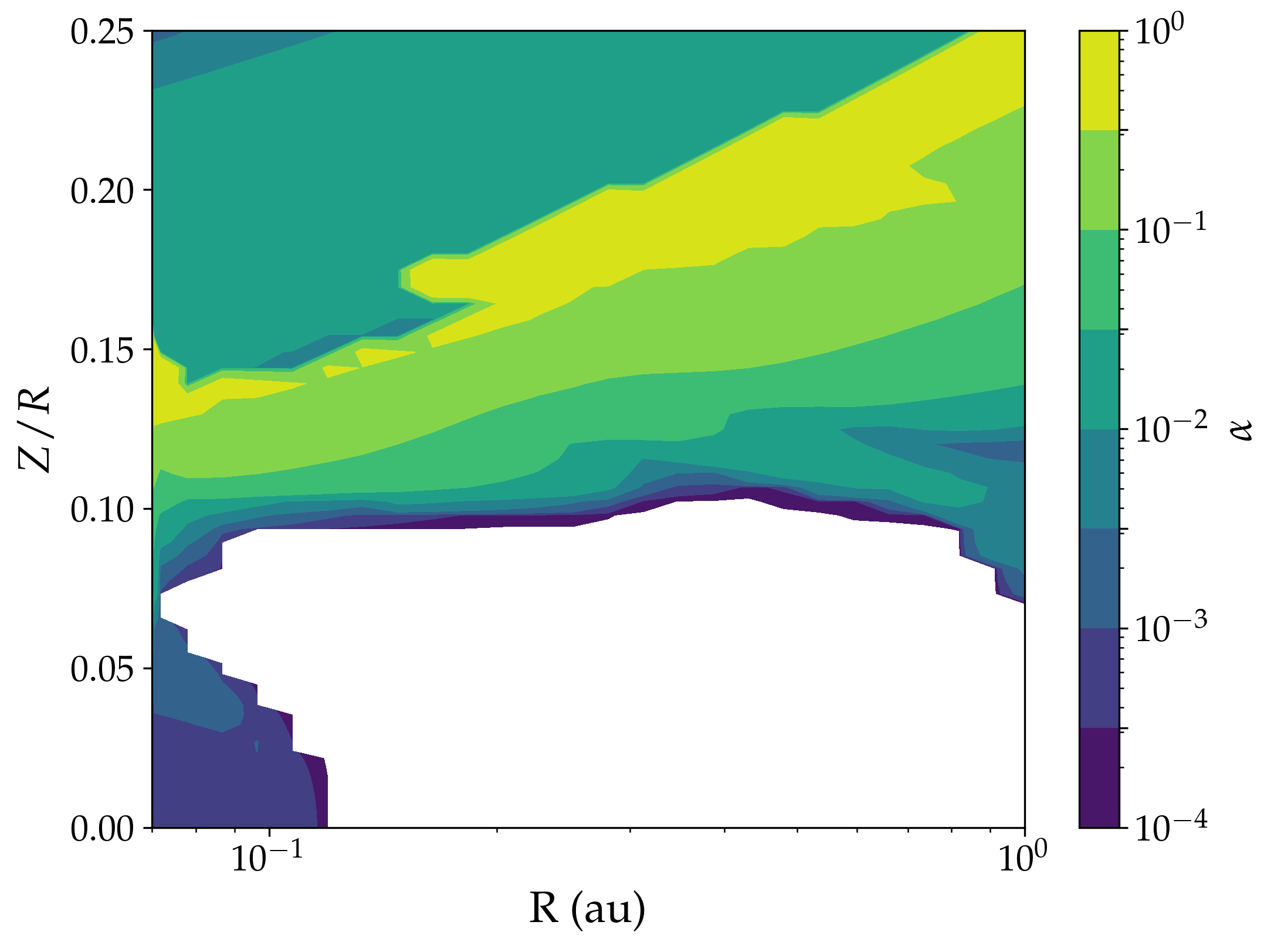}
  \caption{No Flare}
  \label{fig:NonIdealViscosityNoFlare}
\end{subfigure}%
\begin{subfigure}{.3\textwidth}
  \centering
  \includegraphics[width=1\linewidth]{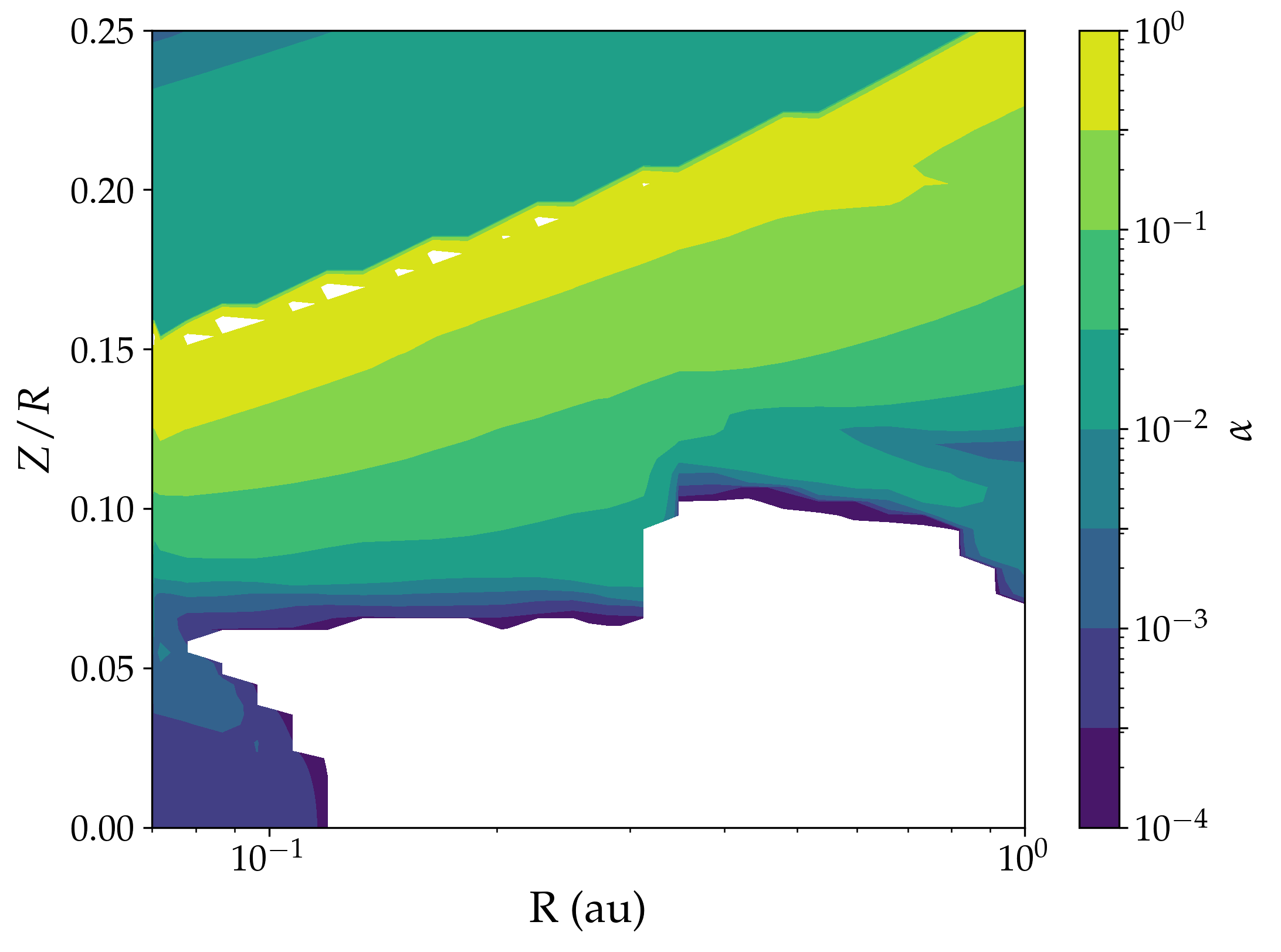}
  \caption{SSF}
  \label{fig:NonIdealViscosityCoronal}
\end{subfigure}
\begin{subfigure}{.3\textwidth}
  \centering
  \includegraphics[width=1\linewidth]{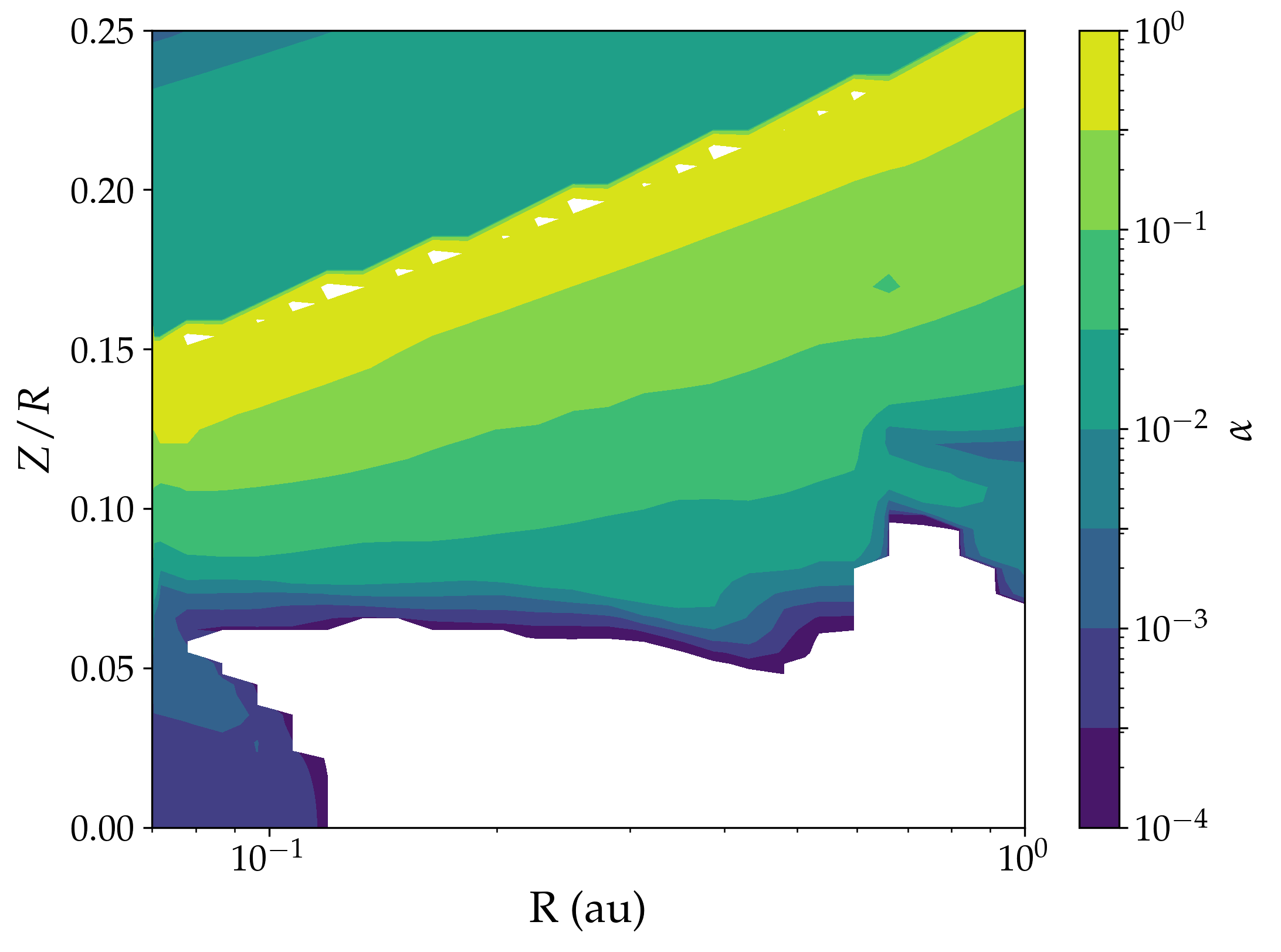}
  \caption{SDF}
  \label{fig:NonIdealViscosityStarDisc}
\end{subfigure}%

\caption{Expected distribution of the the effective viscosity parameter (Eq. \ref{eq:effectiveviscosity}) as a function of the disc radius, $R$, and the normalised height, $Z/R$, in the case of no flare (panel a), SSF (panel b), and SDF (panel c). The white area delimits the MRI inactive region, inside which $\sqrt{\beta_{\rm mag}} \Lambda_{\rm Ohm}<1$.}
\label{fig:NonIdealViscosity}
\end{figure*}

Figure \ref{fig:NonIdealViscosity} displays the distribution of the effective viscosity parameter \( \alpha \) for a \( \beta_{\text{mid}} \) value of \( 10^4 \). This parameter traces the turbulence efficiency, see Sec. \ref{sec:Viscosity}. Three zones can be distinguished, the MRI-active region with high values of $\alpha$, the Ohmic diffusion-limited "dead zone" delimited by the total damping criterion \( \beta_{\text{mag}}^{1/2} \Lambda_{\text{Ohm}} < 1 \), and the portion of the disc where \( \beta_{\text{mag}} < 1 \) in the disc atmospheres leading to a drop of $\alpha<0.01$. The decrease in \( \alpha \), as one moves outward the disc, is linked to the reduction in gas pressure and consequently, \( \beta_{\text{mag}} \). 

In regions with higher ionisation rates, the coupling between the gas and the magnetic field is stronger, leading to more efficient MRI-driven turbulence and angular momentum transport, see Sec. \ref{sec:Viscosity} and \citet{2007ApJ...659..729T}. 

We focus now on the effects of additional ionisation due to flares. In the region undergoing an increased ionisation rate, at $R < R_i$, where $R_i=0.3$ au and $R_i=0.6$ au for SSF and SDF, respectively, the values of the viscosity parameters increase from 0 to a finite value $\gtrsim 0.01$, activating MRI. On Fig. \ref{fig:NonIdealViscosity} panels (b) and (c), we see that the size of the MRI active region increase from $Z/R\approx0.1$ to $Z/R\approx 0.06 $. In this interval, the MRI active region size increases by more than one order of magnitude in terms of column density. 

\begin{figure}
    \centering
    \includegraphics[width=1\linewidth]{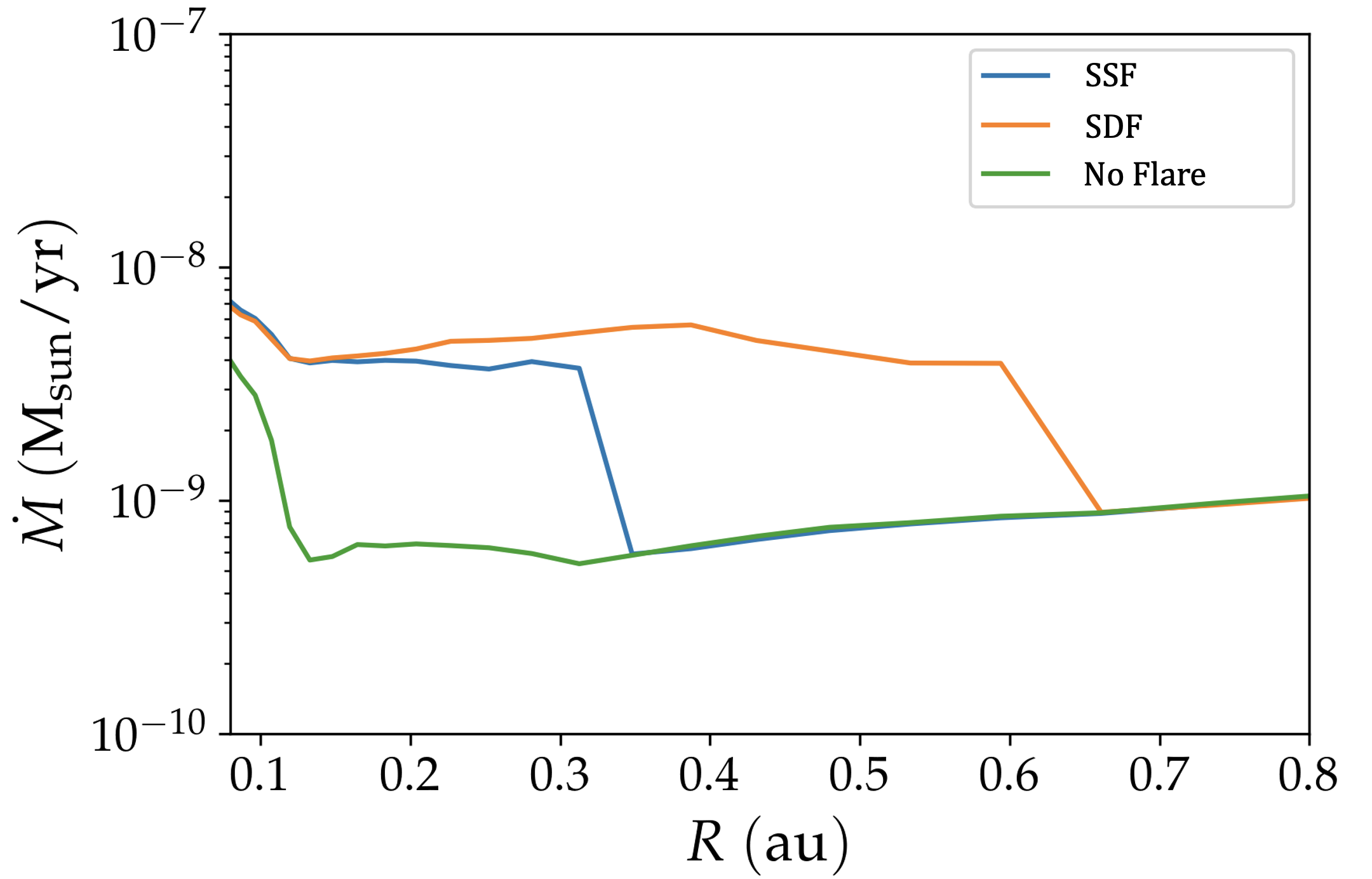}
    \caption{Accretion rate in the inner disc assuming a disc model impacted by the EPs produced by SDF (orange), SSF (blue) and in the absence of flare (green).}
    \label{fig:accretionrates}
\end{figure}

By controlling the disc viscosity, ionisation processes regulate the mass accretion rate onto the central star. If the mass accretion is driven by turbulence originating from either hydrodynamical or magnetohydrodynamical processes, assuming efficient angular momentum transport via the MRI, the disc mass accretion rate can be approximated in terms of an $\alpha$-disc model \citep{2011ApJ...735....8P},

\begin{align}
\dot{M} &=  3\pi \nu \Sigma = 2\times 3\pi \mu m_p N_g \alpha c_s^2 \\
& = 2.5 \times 10^{-9} \left(\frac{\alpha}{0.01}\right) \left(\frac{N_g}{10^{24} ~ \rm cm^{-2}}\right) \left(\frac{T}{10^{3} ~ \rm K}\right) ~ \rm  M_\odot/yr,
\end{align}
where the effective viscosity is given by $\nu = \alpha c_s h$ and $\mu$ is the mean molecular weight. $c_s$ and $h$ are the sound speed and the height scale of the disc, respectively, defined in Sec. \ref{sec:Viscosity}. The factor of 2 accounts for accretion along the upper and lower disc surfaces, $N_g$ is the gas column density above the midplane and $T$ is the disc temperature. 
This formula is an approximation in our case as it assumes very efficient angular momentum transport, namely $l_{\rm in}\ll l$ where $l=R^2\Omega$ is the specific angular momentum. This is correct far from the inner disc radius, but it may not necessarily be the case here. Nevertheless, the formula provides an order of magnitude estimation, which requires to be further tested using numerical simulations, like in \citet{2011MNRAS.415.3380O}. In the region undergoing the increased ionisation rate, i.e. at $R < R_i$, where $R_i=0.3$ au  and $R_i=0.6$ au for SSF and SDF, respectively, over 1000 yr, the averaged accretion rate increases by a factor of 7 to 10 compared to the model without flares, see Fig. \ref{fig:accretionrates}. 

\subsubsection{Disc heating by EPs }

Ionisation processes in T Tauri discs contribute to the heating of the disc material, which affects the disc thermal structure \citep{1998ApJ...500..411D,2009ApJ...699.1639E}. Various ionisation mechanisms, including cosmic rays, X-rays, and UV radiation, deposit energy into the disc gas, which in turn raises the gas temperature and modifies the thermal equilibrium \citep{1999ApJ...518..848I,Rab17}.
In this section, we compute the volumetric heating rate due to the ionisation of EPs produced by flares. 

\begin{figure*}
\begin{subfigure}{.48\textwidth}
    \includegraphics[width=1\linewidth]{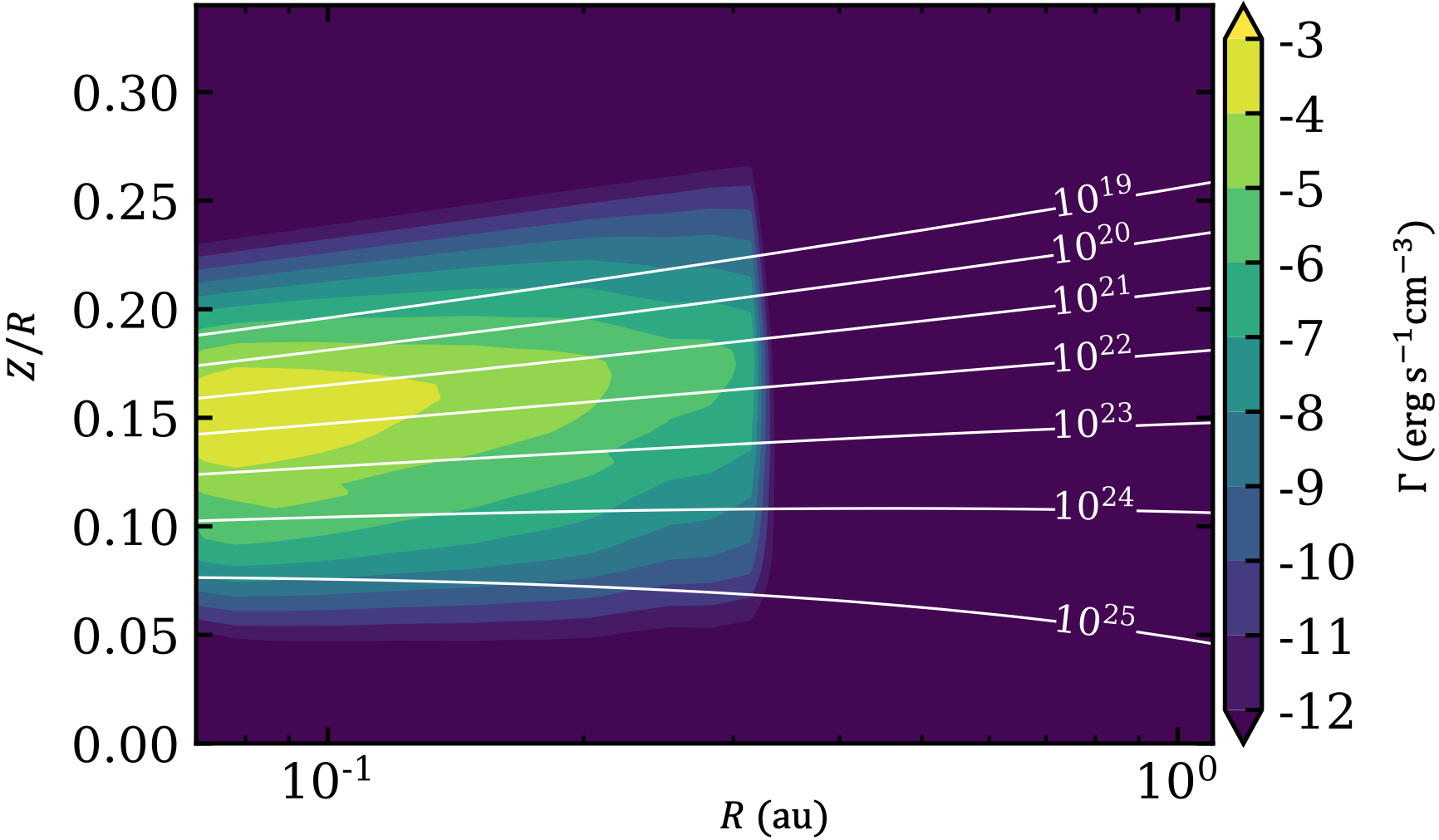}
    \caption{SSF}
\end{subfigure}
\begin{subfigure}{.48\textwidth}
    \includegraphics[width=1\linewidth]{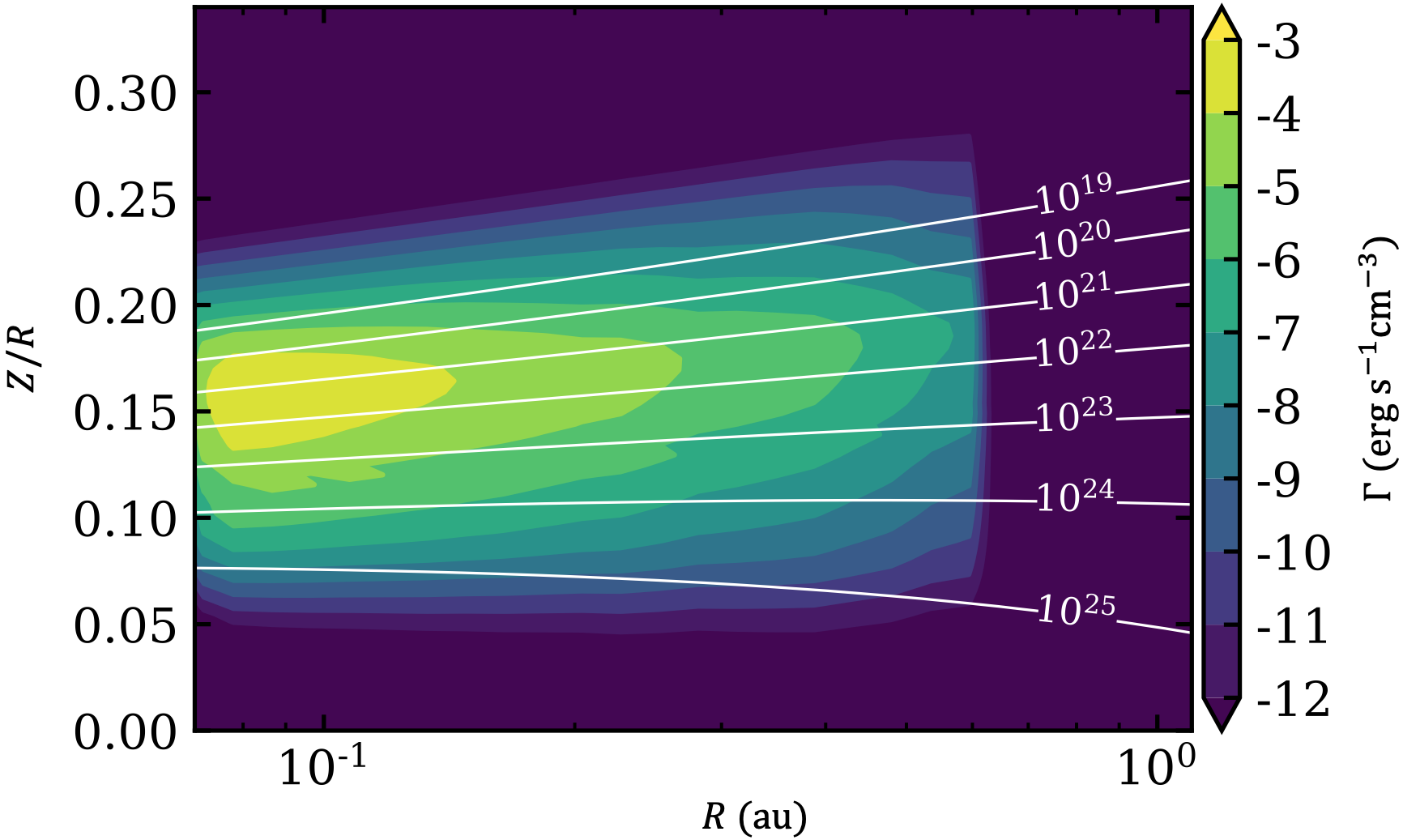}
    \caption{SDF}
\end{subfigure}
\caption{Expected distribution of the volumetric heating rate as a function of the disc radius, $R$, and the normalised height, $Z/R$, as calculated in \prodimo in the case of SSF (panel a), and SDF (panel b). The white solid contour lines show the total vertical column density.}
\label{fig:HeatingRateProDiMo}
\end{figure*}

When they are separated from their nucleus, electrons deposit heat in the surrounding medium. 
The amount of energy deposited per unit time and unit volume is called the volumetric heating rate. The volumetric heating rate due to EPs, $\Gamma$ is computed assuming that the ionisation by each atomic hydrogen releases an amount of energy $Q_{\rm H}=4.3$ eV and ionisation by each molecular hydrogen releases $Q_{\rm H_2}=18$ eV \citep{Glassgold_2012}. $Q_{\rm H}$ and $Q_{\rm H_2}$ are the sum of the heating produced by elastic collisions, rotational excitation, excitation of H$_2$ vibrational levels, dissociation of H$_2$, and chemical heating produced by the reactions of the H$^+$ and H$_2^+$ ions with other chemical species in the disc. All these heating processes are detailed in \citep{Glassgold_2012}. So, 
\begin{equation}
    \Gamma(R,N)= \left(n_{\rm H} (R,N) Q_{\rm H}+ n_{\rm H_2} (R,N) Q_{\rm H_2}\right) \zeta(R,N),
    \label{eq:heatingrategeneral}
\end{equation}
where $\zeta$ is the ionisation rate per H nuclei.
The volumetric heating rate in the disc as calculated in \prodimo is shown in Fig.\ref{fig:HeatingRateProDiMo}.

Previous works have indicated that heating the surface of the disc can lead to an increase in the mass of the outflow \citep{2000A&A...361.1178C,2016ApJ...818..152B}. \citet{2000A&A...361.1178C} showed that introducing a heating source comparable to emissions from an embedded stellar object could offset magnetic compression, thereby allowing for lower mass loads. Previous works simulating the launching of jets incorporate heating rates at the disc surface. The primary sources of disc ionisation previously considered were high-energy stellar X-ray and UV radiation, which also act as the predominant heating agents for both, the disc surface and its entire wind zone. In the absence of magnetic fields, this level of heating is powerful enough to initiate the photoevaporation of protoplanetary discs. This process is widely considered as the leading mechanism for disc mass loss and dissipation, as highlighted in the review by \citet{2014prpl.conf..475A}.

\begin{figure}
\includegraphics[width=\linewidth]{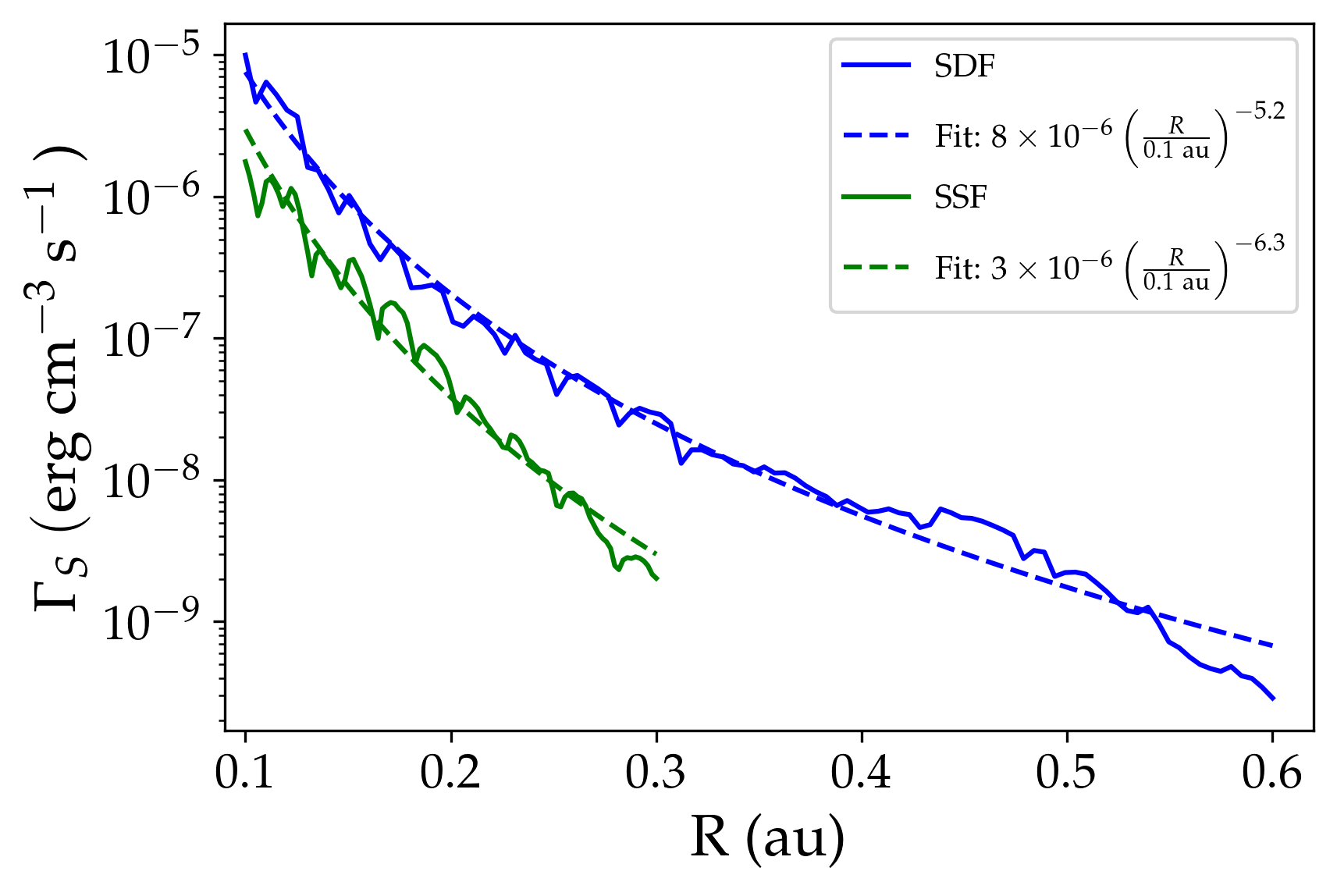}
\caption{Average volumetric heating rate at the surface layer between $10^{19}-10^{20} ~ \rm cm^{-2}$. We consider the SSF model (green) and the SDF model (blue). Solid lines plot the fits for the volumetric heating rate (Eq. \ref{Eq:gammas}).}
\label{fig:SurfaceHeatingRate}
\end{figure}

\begin{figure*}
\includegraphics[width=1\linewidth]{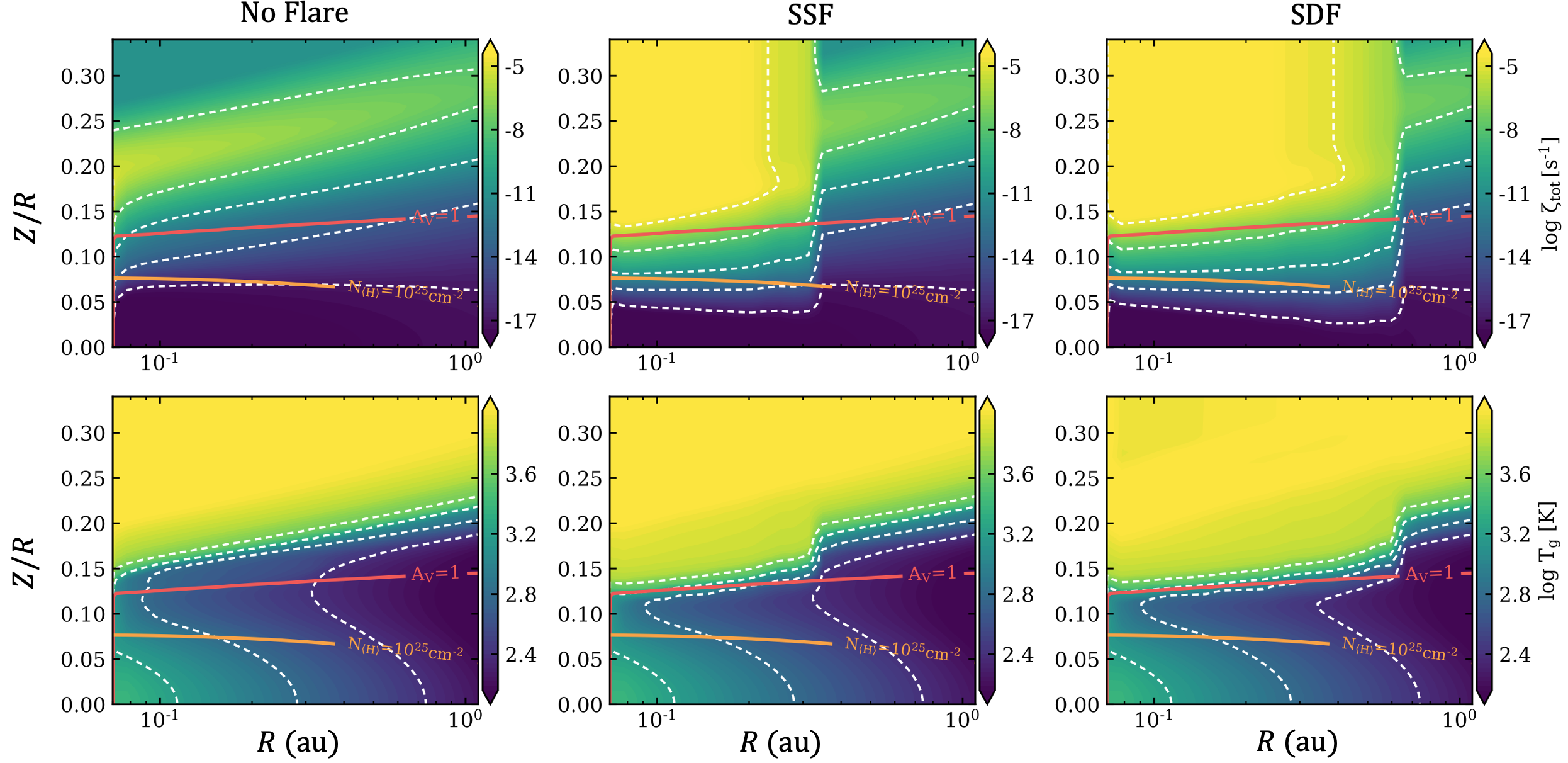}
\caption{From left to right disc models with no flares, SSF and SDF. Top row: Total (X-ray + SP + cosmic rays) ionisation rate per molecular hydrogen. Bottom panel: Gas temperature. The red solid contour shows the visual extinction of unity, and the orange solid contour shows the total vertical column density of $10^{25}\,\mathrm{cm^{-2}}$. }
\label{fig:disciongas}
\end{figure*}

The intensity of this volumetric heating rate influences in turn the jet and the wind characteristics, such as the degree of collimation, velocities, mass loss rates, and temperatures, see Meskini et al. in prep.
In these studies however, the microphysical origin of the volumetric heating rate is not discussed. In the context of the recent advances in the development of thermal-MHD outflow models \citep{wang2019global,gressel2020global,rodenkirch2020global,lesur2021systematic}, we derive a heat source based on microphysical consideration. 
We compute $\Gamma_S$, the average volumetric heating rate in the surface layer, at column densities between $10^{-19}$ and $10^{-20}$ cm$^{-2}$. Figure \ref{fig:SurfaceHeatingRate} shows the trends of $\Gamma_S$, which is parameterised by the power-law,

\begin{equation}
    \Gamma_S\approx a \times 10^{-6}  \left(\frac{R}{0.1 ~  \rm au}\right)^{-b} ~\rm erg ~ cm^{-3} ~s^{-1},
    \label{Eq:gammas}
\end{equation}
where $a= 3$ and $b=6.3$ for the SSF model, while $a=8$ and $b=5.2$ for the SDF model.

The volumetric heating rate computed here can be considered as an upper limit for the volumetric heating rate in outflows. In outflows, the density is lower than the density in the disc. This would lead to lower heating rates in the outflows, see Eq. \eqref{eq:heatingrategeneral}. Thermal MHD simulations show that strong heating rates as the one computed here lead to uncollimated outflows, see Meskini et al. (in prep.).

While MHD winds are generally considered to drive stellar accretion through mass loss, thermal winds have been suggested as a mechanism for dispersing protoplanetary discs. Specifically, these thermal winds, also known as photoevaporative winds, are driven purely by thermal forces and do not contribute to the angular momentum transfer. They were initially proposed by \citet{hollenbach1994photoevaporation} to explain the relatively short lifetimes of discs, of the order of a few million years. In these models, the gas at the surface is heated by photons and forms a wind that is thermally energised. The rotating gas has enough energy to escape from the stellar gravitational potential. Although the launching radius of a thermal wind for a T Tauri star is around $\approx 1\,\mathrm{au}$ \citep[e.g.]{Ercolano2022} there is also a non-zero flow inside the launching radius \citep[e.g.][]{Dullemond2007}
 
We show here that the ionising particles produced by flares are also a source of heating that should be taken into account in the thermal equilibrium of the disc for the launching of magneto-thermal winds \citep[e.g.][]{2016ApJ...818..152B}.

\section{Discussion}\label{S:DISCUSSION}

\subsection{Viscous heating}
The thermal equilibrium in discs should not only integrate chemical and radiative heating and cooling processes but also account for viscous accretion heating $\Gamma_{\rm acc}$. For a thin disc, this last quantity can be expressed as \citep{pringle1981accretion},
\begin{equation}
    \Gamma_{\rm acc}= \frac{9}{4} \alpha \rho c_s^2,
    \label{eq:accretionvisc}
\end{equation}
where $\rho$ is the gas mass density and $c_s$, the sound speed. In this framework, the parameter \( \alpha \) links the viscous heating to the gas pressure ($P_g=\rho c_s^2$). This equation is fundamentally phenomenological and comes with substantial uncertainties. We use it here just to build an illustrative idea of the physical processes at play. 

Disc models that include such a heating source are particularly dependent on the value of \( \alpha \) for determining thermal structure. Former studies, such as those by \citet{1998ApJ...500..411D,d1999accretion,glassgold2004heating}, have investigated the impact of different values of \( \alpha \) on the thermal properties of the gas, commonly based on models with \( \alpha \approx 0.01 \). In the aforementioned models, X-ray heating dominates in the disc corona, while accretion heating is more important in the inner disc regions.

In our study, we propose that EPs from flares could serve as a significant source of heating, both at the surface and within the inner disc, primarily emanating from two mechanisms. Firstly, heating is released through the ionisation of the disc atoms and molecules (see Fig. \ref{fig:HeatingRateProDiMo}). Secondly, these EPs alter the chemical and ionisation structure, thereby increasing the disc viscosity (see Fig. \ref{fig:NonIdealViscosity}). This increased viscosity triggers an increase of the accretion rate, leading to enhanced accretion heating (see Eq. \eqref{eq:accretionvisc}).

These two additional sources of heating could, in turn, alter the disc thermal structure. We will quantify these changes caused by the additional heating sources in an upcoming paper. Meanwhile, we speculate that the temperature increase due to heating by EPs produced by flares may lead to an expansion of the region in which the disc temperature is high enough to maintain sufficient ionisation levels for triggering the MRI.

\subsection{Variability of accretion-ejection processes}

Accretion variability depends on the physical conditions within the disc and the processes that generate unstable flows. Additionally, the amplitude of this variability depends on how efficiently the involved processes can extract energy and angular momentum from the disc. 

For events occurring at a disc radius \( R \) and influenced by gravitational forces, the local Keplerian time is given by,
\begin{equation}
     t_{K} \sim 1 \left(\frac{M_*}{M\odot}\right)^{-1/2} \left(\frac{R}{1 ~\rm au }\right)^{3/2} \rm yr.
\end{equation}
It ranges from just a few days near the inner edge of the disc, to centuries for material in the outer disc. However, if the diffusion of the disc material is the determinant process as in viscous \( \alpha \)-disc models for mass transport, then the diffusion time, $t_{\text{diff}} =\frac{R^2}{\nu} $, is,
\begin{equation}
    t_{\text{diff}} (R) \sim 10^4 \left(\frac{0.01}{\alpha}\right) \left(\frac{0.1}{h/R}\right)^2 t_{K}(R).
\end{equation}
It sets the time scale. Here, \( h \) represents the disc scale height, where typically \( \alpha < 0.01 \) and \( h/R < 0.1 \). Consequently, even in the innermost parts of the disc, viscous diffusion is expected to require a minimum timescale of several years.

Moreover, the thermal timescale defined from the sound speed \( c_s = \alpha^{-1} v_{\rm turb} \) falls between the dynamical and diffusion times, 
\begin{equation}
    t_{\text{th}}(R)\sim 10^2 \left(\frac{0.01}{\alpha}\right) t_{K}(R)
\end{equation}
and corresponds to months near the disc inner edge. 

Considering these two characteristic time scales, we infer that the increase in the accretion rate due to flares can be triggered by a corresponding average increase in ionisation over periods larger or similar to \( t_{\text{diff}} \). Consequently, viscous accretion likely does not play a role in accretion variability over periods shorter than a hundred years in the inner disc. 

On the other hand, flare activity can impact the thermal structure on shorter timescales. Accretion triggered by the heating pulse of a flare, as simulated by \citet{2011MNRAS.415.3380O,2019A&A...624A..50C}, may play a role on the accretion variability on timescale of months, which is the typical timescale of superflares as observed by \citet{Getman_2021a}.

In addition, as the modification of the thermal structure also affects the launching of winds and jets, the variability in flare activity might play a role in the variability of outflows at timescales of months.

\subsection{Effects of EP in jets and winds launching}
In the current understanding, the chemical state and the
thermal energy equilibrium in wind regions is determined by the local flux of stellar high energy UV and X-rays photons. Determining the gas temperature is crucial for accurately calculating pressure gradients, wind flow characteristics, and interpreting line emissions. The gas temperature is influenced by the interplay between various heating and cooling processes. As for the heating, factors like collisions with dust grains, and line emissions from ions, atoms, and molecules come into play. These molecular abundances are also dependent on underlying chemical reactions.

At the surface of the disc, where the wind originates, multiple factors contribute to the gas heating. This includes grain photoelectric heating by FUV photons. It also includes the conversion of binding energy to heating during the formation of molecular hydrogen. Finally, it includes the excess electron energy resulting from photo-dissociation and photo-ionisation by various forms of radiation, i.e. FUV, extreme Ultraviolet and X-rays. We have shown that in this surface layer, EPs produced by flares may be an important additional ionising and heating source.

Above the disc, magnetised jets and winds may also experience heating through non-ideal MHD processes like Alfv\'{e}n wave damping heating, Ohmic heating and ambipolar diffusion heating, especially at altitudes where photo-heating is less effective, as shown by various studies (e.g., \citealt{garcia2001atomic,panoglou2012molecule,wang2019global}). Based on the results presented in this article, which focuses on discs, we emphasise the importance of studying the effects of particles generated by flares also on winds and jets. Such a model will be presented in a forthcoming publication.

\subsection{Dependence of the results to the model parameters}
\subsubsection{Waiting-Time distribution} 

In this section, we explore how the distribution of waiting intervals between two flares affects the ionisation rate. The aim of this analysis is to assess the possible variation in ionisation rates considering the uncertainty associated with the determination of the parameters of this distribution.
To randomly generate the luminosity of the flares, we have developed this distribution using three parameters: the total annual frequency of flares ($N_{\rm tot}$), the annual frequency of super flares ($N_{\rm MF}$), and the power-law index ($a$), see Sec. \ref{sec:waitingtimes}.
The estimation of $N_{\rm tot}$ and $a$ is based on the extrapolation of solar properties, which is a strong assumption and introduces significant uncertainty. Conversely, the estimation of $N_{\rm MF}$ is supported by robust statistics thanks to observations of flares on young stars made by \citet{Getman_2021a,Getman_2021b}, leading to short range of acceptable values.

We note that our model is primarily dependent on the frequency of super flares $N_{\rm MF}$. Because according to our flare geometry model, the loops of these high-luminosity flares are the only long enough to send particles into the disc. However, given the short range of acceptable values for this quantity, its variation has minimal impact on the results. On the other hand, although the uncertainty on the parameters $N_{\rm tot}$ and $a$ is larger, their limited influence on the ionisation rate means this uncertainty does not significantly affect the results. Varying $N_{\rm tot}$ and $a$ with $N_{\rm MF}$ constant is equivalent to varying the flare frequency while keeping the energy deposited by the super flares constant.

In conclusion, varying the three parameters of the waiting time distribution within their uncertainty range does not significantly affect the average ionisation rate.

\subsubsection{Energetic particle density entering the disc}
The most critical parameter in our model is the density of EPs penetrating the disc. This quantity is notably challenging to be constrained by observations. We have outlined a method in B23 based on the assumption that energetic electrons emit the observed X-ray radiation via bremsstrahlung. From these observations, we have estimated the density of the thermal plasma emitting X-rays. Assuming that this density is that of the reconnection region, we normalised the number of supra-thermal particles entering the disc. The uncertainty in our approach is that we are not certain that the density derived from bremsstrahlung radiation corresponds to that of the reconnection zone. It could actually be the density of the stellar or the disc chromosphere heated by the particles. If this is the case, the density that we assumed would be overestimated. We find that if the density of EPs entering the disc is reduced by a factor greater than about 3 orders of magnitude compared to that estimated in B23, the ionisation rate from particles produced by flares would be negligible compared to the ionisation rate by the star X-ray radiation. In their study, \citet{2022Natur.606..674F} have shown the challenges and requisite resolution for determining the density of supra-thermal electrons in solar flares through radio emission observations. This highlights the complexities involved and suggests that the observational determination of supra-thermal particle density in T Tauri flares is not currently feasible. In Sec. \ref{sec:particlemodellimitation}, we propose a numerical approach that could allow for a more accurate estimation of the density of supra-thermal particles penetrating the disc.

\subsection{Model limitations}

\subsubsection{Particle model} \label{sec:particlemodellimitation}
First, there are aspects of the particle propagation model that we have not addressed, especially when it comes to propagation at column densities greater than \(10^{25} \text{cm}^{-2}\). Beyond this column density, the dominant losses for protons are through pion production. During pion production, protons lose a significant portion of their energy and change their pitch angle. Due to these factors, particle transport can no longer be approached using the CSDA framework \citep{Padovani2009}. Diffusive processes need to be considered. For a more detailed discussion, see \citet{Padovani18}.

Second, in our particle model, we must establish what is the fraction of particles accelerated in the reconnection zone of the flare, that penetrate into the disc. This aspect relies on the flare geometry and the magnetic field structure within the star-disc interaction zone. For now, we will restrict ourselves to a parametric study. We scale down the incoming particle flux to estimate the minimum fraction of particles that must enter the disc to significantly impact its chemistry and dynamics. For a quantitative assessment, a test particle distribution would have to be integrated into a reconnection zone of an MHD simulation of the inner disc. This should allow to compute the fraction of particles that penetrate the disc, fall onto the stellar surface, and propagate into the outflows.

Third, we are also aware that the particle acceleration model from B23 is somewhat simple. However, the models and simulations of particle acceleration by magnetic reconnection are an active research field, see \citet{2020LRCA....6....1M,2022NatRP...4..263J} for reviews. The magnetic structure of the star-disc interaction region does not only influence the location and proportion of particles entering the disc but also, as detailed in B23, the particle acceleration process itself. As shown in B23, the presence of a guide field, the magnetic field component that is not involved in the reconnection process, diminishes the particle acceleration efficiency. We have not accounted for a guide field here, placing us in an optimal configuration for particle acceleration. In a model with a more realistic magnetic structure, featuring a toroidal component that could act as a guide field, the acceleration process would be less effective, thus resulting in lower ionisation rates. 

While the literature discusses the possibility that superflares are sufficiently long to connect the star with the disc \citep{2005ApJS..160..469F,2012ApJ...754...32H} or not \citep{getman2008a,Getman_2021a}, we highlight the potential for particles to be produced by magnetic reconnection events resulting from Keplerian shearing. Indeed, \citet{2023ApJ...954...15F} conducted simulations to model the occurrence of magnetic reconnection events within circumstellar discs, induced by Keplerian shear. The simulations show the generation of turbulence within the magnetic truncation region, leading to flaring phenomena. The resultant energy and luminosity distributions are characterised by power-law behaviours. This is what we postulate in our work based on observations of soft X-rays. This is now corroborated through numerical support. These distributions exhibit a universal character, consistent under various conditions such as the nature of the reconnection mechanisms and the boundary conditions implemented. Such universality leads \citet{2023ApJ...954...15F} to draw analogies between magnetic reconnection processes in inner circumstellar discs and solar flares, a significant assumption of our study.

To summarise, the particle model used here provides an upper limit to the impact of particles on discs. A model that describes large-scale particle acceleration by magnetic reconnection in the specific magnetic environment of T Tauri stars is currently not available. Nevertheless, such a model would be crucial for a complete understanding of the effects of particles produced by magnetic reconnection on discs. For a more realistic picture, a first step could be to add particles in a MHD code in the test-particle limit or beyond, i.e. including a back reaction term.

\subsubsection{Disc model}

A crucial point for interpreting our results is that all our conclusions about the impact of particles are specific to the chemical structure computed by \prodimo. 
For a more complete picture, it would be interesting to examine the impact of the particles by varying the structure of \prodimo discs. For instance, we anticipate that the particle impact would depend on the radial and vertical structure of the disc mass, and this dependence will be evaluated in a forthcoming paper.  

In addition, we would like to caution about the increase in the density of HCNH$^+$ and HCN due to the additional ionisation rate from particles. The \prodimo code, although it considers hundreds of chemical species and thousands of reactions, does not take into account molecules with more than four atoms, except for methanol. For details, we refer to \citealt{kamp2017consistent}). Therefore, the rise in the abundance of HCNH$^+$ might not be physical but rather a dead-end in the reaction network calculated by \prodimo. The recombination of HCNH$^+$ would then lead to an increase in HCN, which might be overestimated. To assess the increase of the abundance of HCN, we would need to conduct a more detailed analysis of the chemical reactions leading to this increase. Subsequently, we could assess the influence of particles from flares in a chemical model that considers more complex molecules with additional atoms. Nevertheless, we can highlight that our conclusions on the increasing complexity of molecules remains valid. Even if the rise in HCN abundance might be due to a chemical reaction dead-end in the network, the chemistry of more complex species will surely be boosted anyway if energetic particles from flares are included in the model.

Another limitation concerns the calculation of the viscous parameter \(\alpha\) inside the disc. Although \(\alpha\) was initially treated as a parameter to study accretion processes in discs, this parameter can be physically deduced from the stress tensor of the Navier-Stokes' equations. The physics underlying this parameter is thus inherently of MHD origin. In this paper, we computed the value of this parameter based on the properties of a static disc. Hence, the derived \(\alpha\) value is inevitably imperfect. The intent of this computation is more a proof of concept, illustrating that the size of a disc dead zone is reduced when considering ionisation by flares. From the \(\alpha\) calculation, we have also estimated the mass accretion rate, taking ionisation due to flares into account. This estimation too, should be viewed as a proof of concepts, given that it relies on the thin disc model of \citet{Shakura73}, applicable far from the disc inner edge, a criterion not necessarily met in our study.

\section{Conclusions}\label{S:CONCLUSION}

The ionisation of protoplanetary discs is crucial for the triggering of MHD instabilities, responsible for the accretion of matter onto the central star and is crucial in the disc chemistry, key to understand the formation of planets and the synthesis of the building blocks of life. 

In the first paper of this series \citep{brunn2023ionization}, we examined the effects of an additional ionisation source based on microphysical and observational considerations. A stationary state of the particle emission was assumed leading to an overestimation of the ionisation rate. In this paper, we address this issue by incorporating time-dependent components into our model.

We started by examining the average ionisation rate generated by a flare to compare it to the ionisation rate from the stellar X-rays. We conclude that both sources produce similar average ionisation rates at low column densities (\(N < 10^{22} \text{cm}^{-2}\)). However, particles generated by the flares produce an ionisation rate between 1 and 3 orders of magnitude higher than the ionisation rate from X-rays at high column densities (\(N > 10^{22} \text{cm}^{-2}\)). 

We used a Monte Carlo analysis to determine the spatial distribution of a temporally averaged ionisation rate. To facilitate the reuse of our numerical results, we proposed a parametric expression for the spatial distribution of the ionisation rate produced by flares (Eq. \ref{eq:ionizationrateparametrization}). 

Based on the thermo-chemical disc structures computed with \prodimo, accounting for energetic particles produced by flares, we studied the impact of this additional ionisation source on the chemistry, viscosity, accretion rate, and volumetric heating rate of protoplanetary discs.

\textit{Impact on Chemistry:} The ionisation fraction increases by at least an order of magnitude up to \(N\sim 10^{25} \text{cm}^{-2}\) in disc region where particles penetrate. The distribution of cations is also significantly altered, especially the density of H\(^+\) increases in the photodissociation region. Additionally, taking into account particles from flares leads to the formation of a layer where HCNH\(^+\) is the most abundant cation at \(Z/R\approx0.12\). The recombination of this cation leads to an increase of a factor of 2 to 3 in the column density of hydrogen cyanide, HCN. This molecule is crucial in prebiotic chemistry to understand the synthetisation of amino-acids, the building blocks of life.

\textit{Impact on Viscosity and Accretion:} 
The extent of the MRI-stable region or "dead zone", is reduced. We estimated that the mass accretion rate increases by almost one order of magnitude in disc regions where the energetic particles produced by flares penetrate.

\textit{Impact on the Volumetric Heating Rate:} 
We determined the volumetric heating rate resulting from the ionisation of disc matter by energetic particles produced by flares. We showed that this volumetric heating rate is strong and would lead to collimated outflows. To facilitate the use of our numerical results, we proposed a parametric expression for the volumetric heating rate at the disc surface (Eq. \ref{Eq:gammas}).

Given that particles from flares affect both the chemical and the thermal structure of the disc in regions that JWST/MIRI is capable of tracing, it is reasonable to expect that these flares would impact JWST spectra. We plan to refine our flare model using the synthetic spectra production feature of \prodimo. However, multiple aspects related to synthetic JWST spectra production are currently under development within \prodimo. So, we postpone this study to a forthcoming publication.

We have shown that EPs produced by flares have strong impact on the inner disc of T Tauri discs. In a forthcoming work, we will present a particle acceleration model due to magnetic reconnection that could occur further out, at the surface of the discs. This work will extend the study presented in this paper to study the impact of energetic particles produced by turbulence-induced reconnection events. 

Additionally, until now, our focus has been on the influence of energetic particles on the ionisation rates, which subsequently affect the disc dynamics and chemistry. However, energetic particles can also collide with stable isotopes, creating radioactive isotopes, such as $p + ^{25}$Al$ \rightarrow ^{26}$Al. The increased abundance of these radioactive species acts as a heat source within the disc. Previous work (e.g \citealt{2020ApJ...898...79G}) has shown that planet formation is significantly influenced by this heating source. Although the formation of planets was not the research topic of this paper, our particle acceleration model could be applied to this broader subject.

\section*{Acknowledgements}
We acknowledge financial support from "Programme National de Physique Stellaire" (PNPS) and from "Programme National des Hautes Energies" (PNHE) of CNRS/INSU, France. This work has been conducted under the INTERCOS project of CNRS/IN2P3. CS thanks LUPM for hosting him  (d\'el\'egations CNRS). Ch. Rab is grateful for support from the Max Planck Society and acknowledges funding by the Deutsche Forschungsgemeinschaft (DFG, German Research Foundation) - 325594231. The authors thank A. Vidotto and P. Caselli, Z. Meliani, S. Masson, T.P. Downes, J. Ferreira, G. Lesur, J. Morin, V. Guillet, D. Rodgers-Lee, for fruitful discussions. 


\section{Data Availability}
The data underlying this article will be shared on reasonable request to the corresponding author.



\bibliographystyle{mnras}
\bibliography{main} 




\appendix

\section{Normalising the particle injection flux - Revision of B23 }\label{app:Nonthermaldensitynormalisation}

We present here a revision of the calculation of B23 of the non-thermal particle flux produced by the flare. 

The energy distribution per unit volume, denoted as $F(E)$, or the flux, denoted as $j(E)$, is expressed in terms of the kinetic energy, $E$, for both electrons and protons, where 
\begin{equation}
    j(E)=F(E) \frac{v(E)}{4\pi},
\end{equation}
and $v$ is the particle speed. The flux and distribution consist of a thermal and a non-thermal component, denoted by the subscript "th" and "nt", respectively. It is expressed in units of particles per unit energy, time, area, and solid angle.

The thermal component is defined by its temperature $T$ and its normalisation $n_{\rm th}$, where $n_{\rm th}$ is the thermal particle density in cm$^{-3}$. For energetic protons and electrons we assume equal temperatures. This gives us the following equation for their energy distribution,

\begin{equation}
\label{jTH}
  F_{\rm {th}}(E) =  n_{\rm th} \frac{2}{\sqrt{\pi}} \sqrt{\frac{E}{kT}}\frac{1}{k T} \exp(-E/kT) \ .
\end{equation}
Here, the peak value is at $E = 3/2 k T \equiv E_{\rm th}$.

As it is usually assumed in reconnection models for solar flares  \citep{ripperda2017reconnection} and supported by observation of solar flares \citep{emslie2012global,matthews2021high} energy equipartition between non-thermal protons and electrons below 1 GeV is a good proxy of non-thermal content energetics. The energy density of non-thermal particles can be represented as,

\begin{equation}
U_{e,\rm{nt}}= U_{p,\rm{nt}}=\int_{E_{ c}}^\infty E F_p(E) \dd E=\int_{E_{ c}}^\infty E F_e(E) \dd E
\ ,    
\end{equation}

If we assume that electrons and protons have the same injection energy $E_{ c}$, they also share the same non-thermal energy distribution, 
\begin{equation}
    F_e(E)=F_p(E)=F_{\rm nt}(E).
    \label{eq:equipartionofenergy}
\end{equation} 

As a result of the magnetic reconnection process, we introduce the non-thermal component of the energy distribution function $F_{\rm nt}$. The non-thermal energy distribution is determined by its normalisation $N_{\rm nt}$. In the following, we show a way to derive this quantity.

We define a maximum energy for the non-thermal distribution, $E_{ U} > E_{ c}$. The distribution is a power-law between these two energies. The index of the power law is called $\delta$. This model is consistent with the non-thermal electron distribution observed in solar flares. 

The injection energy, $E_{ c}$, is proportional to the thermal energy, $E_c=\theta E_{\rm th}$. Most T Tauri flare models exhibit a single power law. We assume an exponential cut-off beyond $E_{ U}$,

\begin{equation}
\label{FNTH}
    F_{\rm nt}(E)=N_{\rm nt} \left(\frac{E}{E_{ c}}\right)^{-\delta} \exp\left(-\frac{E}{E_{ U}}\right),
\end{equation}

The normalisation factor $N_{\rm nt}$ can be expressed in terms of the non-thermal particle density,
\begin{equation}
    n_{\rm nt}=\int_{E_{ c}}^\infty \left(\frac{E}{Ec}\right)^{-\delta}\exp\left(-\frac{E}{E_{ U}}\right) dE \simeq N_{\rm nt} \frac{E_c}{\delta-1}.
\end{equation}

The total electron density as function of the flare temperature $T$ derived from X-ray observation (B23), 
\begin{equation}
n_{e}(T)=20.5 \times 10^{10} \left(\frac{T}{1 ~ \rm MK}\right)^{-0.69} ~ \rm cm^{-3}
\label{eq:obsdensity}
\end{equation}

is the sum of the thermal and non-thermal components, $n= n_{\rm th}+ n_{\rm nt}$. By setting the non-thermal and the thermal energy distribution to be equal at the injection energy ($E=E_c=\theta E_{\rm th}$), we can derive $n_{\rm nt}$,

\begin{equation}
    n_{\rm nt}=\frac{3\theta}{\delta-1}\sqrt{\frac{3\theta}{2 \pi}} \exp\left(-\frac{3}{2}\theta\right) n_{\rm th}.
    \label{eq:nonthermaldensity}
\end{equation}
The non-thermal particle density depends on the local density of the thermal plasma. We use the total plasma density $n$ from X-ray observations given by Eq. \eqref{eq:obsdensity} to deduce the thermal and non-thermal densities.
Given that $n=n_{\rm nt}+n_{\rm th}$, we get,

\begin{equation}
    n=n_{\rm th}\left(1+\frac{3\theta}{\delta-1}\sqrt{\frac{3\theta}{2 \pi}} \exp\left(-\frac{3}{2}\theta\right)\right)\ .
    \label{eq:totaldensity}
\end{equation}

From Eqs. \eqref{eq:totaldensity} and \eqref{eq:obsdensity}, the thermal and non-thermal particle densities can be expressed as function of the observed temperature,
\begin{equation}
    n_{\rm th}= 20.5 \times 10^{10} \frac{\left(\frac{T}{1 ~ \rm MK}\right)^{-0.69}}{1+\frac{3\theta}{\delta-1}\sqrt{\frac{3\theta}{2 \pi}} \exp\left(-\frac{3\theta}{2}\right)} \quad \rm cm^{-3},
    \label{eq:thermaldensity}
\end{equation}
Assuming the fiducial values $\theta=3$ and $\delta=3$,
\begin{equation}
    n_{\rm th}=19 \times 10^{10} \left(\frac{T}{1 ~ \rm MK}\right)^{-0.69} ~ \rm cm^{-3}.
    \label{eq:thermaldensityfiducial}
\end{equation}

The non-thermal particle density $n_{\rm nt}$ is then recovered from Eqs. \eqref{eq:thermaldensity} and \eqref{eq:nonthermaldensity},
\begin{equation}
    n_{\rm nt}=20.5 \times 10^{10} \frac{\left(\frac{T}{1 ~ \rm MK}\right)^{-0.69}}{1+\frac{\delta-1}{3}\sqrt{\frac{2\pi}{3 \theta}} \exp\left(\frac{3}{2\theta}\right)} \quad \rm cm^{-3}
    \label{eq:nonthermalfluxnormaisation}
\end{equation}
Assuming the fiducial values $\theta=3$ and $\delta=3$,
\begin{equation}
    n_{\rm nt}= 10^{10} \left(\frac{T}{1 ~ \rm MK}\right)^{-0.69} ~ \rm cm^{-3}. 
    \label{eq:nonthermaldensityfiducial}
\end{equation}
  
We highlight here the difference between Eq. \eqref{eq:totaldensity} and Eq. 15 of B23. In that article, Eq. (15) is non-homogeneous. There is obviously a confusion between the normalisation factor of the non-thermal flux and the non-thermal particle density. We corrected the issue in this section. 
The wrong estimation of the normalisation factor $N_{\rm B23}$ lead to an overestimation of the flux in B23. We compare the normalisation $N_{\rm B23}$ with the corrected normalisation factor $N_{\rm nt}$,
\begin{equation}
    \frac{N_{\rm B23}}{N_{\rm nt}}= \frac{E_c}{k T (\delta-1)}= \frac{3\theta}{2(\delta-1)}.
\end{equation}

Taking the fiducial values of B23, $\delta=3$ and $\theta=3$, $N_{\rm B23}\approx 2 N_{\rm nt}$, the fluxes and ionisation rates are reduced by a factor of two compared to B23.


\section{Collisional Ionisation }\label{app:collion}
In the chemical network used in \citet{Rab17}, collisional ionisation, which can dominate the ionisation fraction in the innermost high density and high temperature ($T > 1000$ K) \citep[e.g.][]{fromang2002ionization,Desch2015} region of the disc, is not included. To model this process, which is usually described by the Saha equation, we add the following reaction to our network. 

\begin{equation}
    \mathrm{H_2} + \mathrm{Na} \rightarrow \mathrm{H_2} + \mathrm{Na}^+ + \mathrm{e^{-}}. 
\end{equation}
Usually, gas phase potassium $\mathrm{K}$ is used because of its low ionisation potential \citep[see][]{Desch2015}. However, $\mathrm{K}$ is not included in our chemical network, and including it would require adding a couple of other chemical reactions (e.g. recombination). We therefore use the element $\mathrm{Na}$ as it is already in our network, but assume the ionisation potential of $\mathrm{K}$ to calculate the rate coefficient using Eq.~6 of \citet{Desch2015}. 

\begin{figure}
    \centering
    \includegraphics[width=1\linewidth]{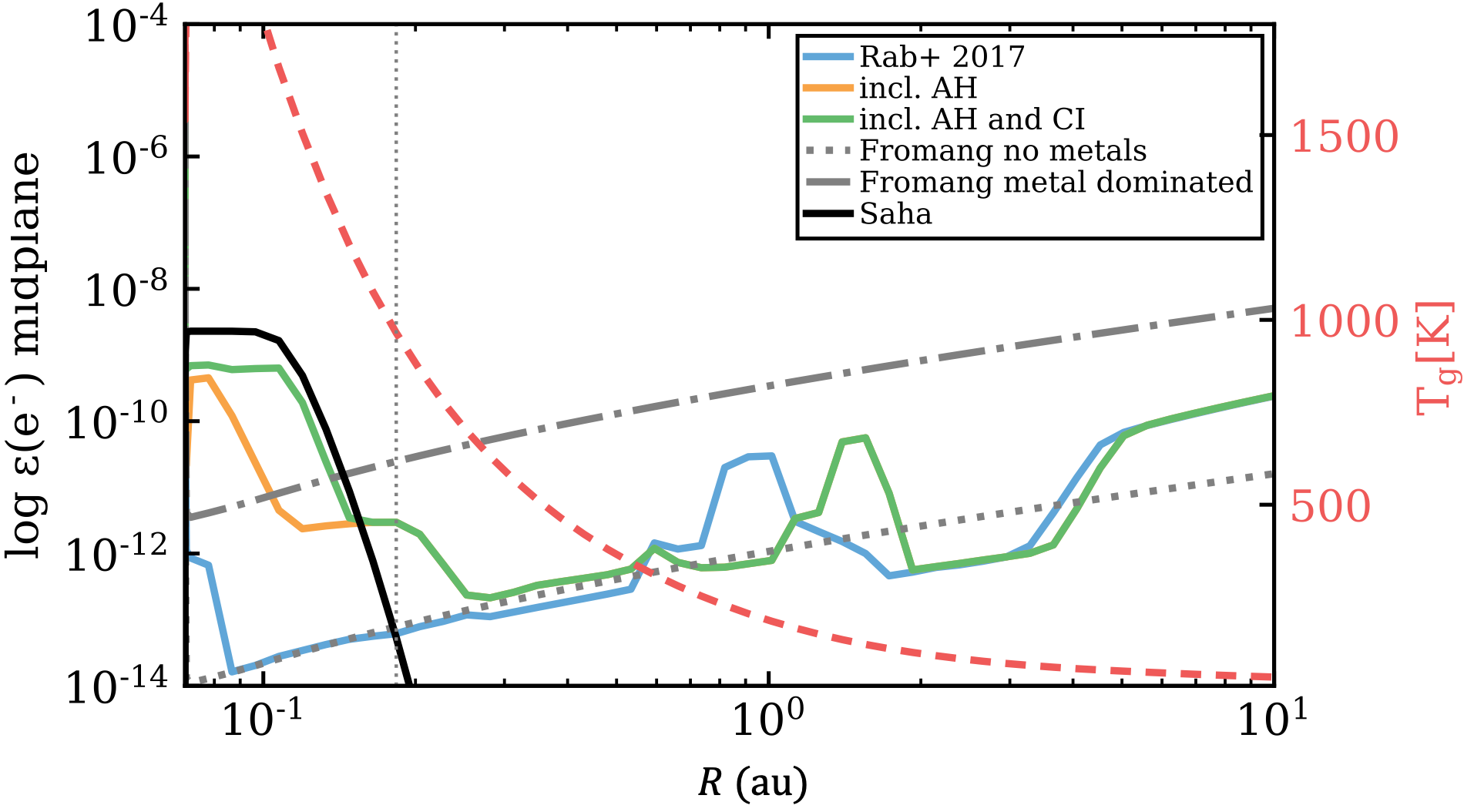}
    \caption{Comparison of the ionisation fraction for different models and (semi-)analytical expressions in the disc midplane. The coloured solid lines show the different thermo-chemical models. In blue, the model from \citet{Rab17}, in green the same model but including accretion heating and in green including accretion heating (AH) and our approximation for collisional ionisation (CI). The red dashed line shows the midplane gas temperature considering accretion heating. The dotted and dash-dotted grey lines show the ionisation fraction using Eq.~13 of \citet{fromang2002ionization} for the no-metals and metal-dominated cases. The solid black line shows the results of solving the Saha equation. The vertical dotted line indicates $T=1000\,\mathrm{K}$.}
    \label{fig:colion}
\end{figure}

To verify our approximation, we compare our models to various analytical expressions in Fig.~\ref{fig:colion}. By simply including accretion heating, which increases the temperature in the midplane, the ionisation fraction increases by two orders of magnitude at $R\approx 0.1\,\mathrm{au}$. The inclusion of collisional ionisation increase by two orders of magnitude and expands the high ionisation region radially. This last model is in good agreement with the ionisation fraction solely considering the Saha equation (using $\mathrm{Na}$ as the main element with the ionisation potential of K). Close to the inner radius, our approximation under-predicts the ionisation rate compared to the Saha equation. This might be because our chemical network also considers other elements and molecules, which are neglected in the Saha equation. Nevertheless, with the addition of the accretion heating and collisional ionisation, the disc regions with $T=1000\,\mathrm{K}$ now have a high enough ionisation rate to become MRI active, which was not the case in the model of \citet{Rab17} (see Sec. \ref{sec:feedbackMRI}). 

\bsp	
\label{lastpage}
\end{document}